%% file: torus-topological-entropy.tex
\documentclass[11pt,a4paper]{article}
\usepackage{jheppub}
\makeatletter
\renewcommand{\@fpheader}{}
\makeatother
\usepackage[T1]{fontenc}
\usepackage{lmodern}
\usepackage{mathtools,amssymb,amsthm,mathrsfs,microtype}
\newtheorem{theorem}{Theorem}[section]

\newtheorem{corollary}[theorem]{Corollary}
\title{Strong subadditivity of topological entropy on the torus: R\'enyi order and non-Abelian structure}
\author[1]{Yi Liang\,\href{https://orcid.org/0009-0006-3168-972X}{\includegraphics[height=1.5ex]{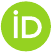}}\note{ORCID: \href{https://orcid.org/0009-0006-3168-972X}{0009-0006-3168-972X}.}}
\affiliation{Independent Researcher, Beijing, China}
\emailAdd{iantrans2042@gmail.com}
\abstract{\input{sections/abstract.tex}}
\hypersetup{pdfauthor={Yi Liang},pdftitle={Strong subadditivity of topological entropy on the torus: Renyi order and non-Abelian structure}}
\begin{document}
\maketitle
\section{Introduction}\label{sec:introduction}
\input{sections/introduction.tex}
\input{sections/objects.tex}
\input{sections/mechanism.tex}
\input{sections/positive.tex}
\input{sections/microscopic.tex}
\input{sections/deformation.tex}
\section{Discussion}\label{sec:discussion}
\input{sections/discussion.tex}
\subsection*{Data availability}
The analytical derivations and region constructions are given in this article.
\appendix
\input{sections/appendix-weighted.tex}
\bibliographystyle{JHEP}
\bibliography{references}
\end{document}

%% file: sections/abstract.tex
We classify universal strong subadditivity of the canonical topological entropy on an unpunctured torus across all pure logical states and admissible region pairs, for finite unitary modular tensor categories and finite positive R\'enyi orders. Below order one, universal validity is equivalent to Abelian, or pointed, anyon structure. At the von Neumann order it holds for every category. In nonpointed categories, shared topological flux and unequal occupied quantum dimensions curve the multi-annulus response, producing violations at every finite positive order other than one. Every category of rank greater than one admits a single state and region pair that violate the inequality for all finite orders at least two. The remaining pointed orders strictly between one and two are governed by an exact equal-order modular Fourier criterion. On the lattice, a specified quantum-double reference ground state for each finite non-Abelian group realizes the nonpointed violations directly in ordinary physical-edge entropies on a common refinement. Applying the same diagonal filter to every physical edge gives a controlled deformation of this reference state. For each fixed group, a common size-independent small parameter interval supports a uniform R\'enyi-two deficit bound and a uniformly gapped local parent for all cost profiles bounded between zero and one, within the stated lattice family. Nonconstant profiles change regional entanglement spectra near the fixed point at every fixed finite volume in the specified strip geometry. The deficit converges to its strictly negative fixed-point value in the controlled joint limit of system size and interface separation.

%% file: sections/introduction.tex
In gapped two-dimensional systems, a universal contribution to ground-state entanglement reveals topological information beyond local order parameters~\cite{KitaevPreskill2006,LevinWen2006}. Spatial entanglement is simultaneously constrained by general information-theoretic inequalities, foremost among them strong subadditivity at the von Neumann order~\cite{LiebRuskai1973}. Moving away from order one by varying the R\'enyi index reweights the underlying entanglement spectrum~\cite{LiHaldane2008}. We ask whether topological order suffices to protect a spatial entropy inequality across this reweighting, and which algebraic properties of the anyon theory govern its survival.

Noncontractible cuts on a torus access the logical flux of the ground state. Across surgery, edge-theoretic, and anyonic formulations, the canonical topological contribution is governed by boundary topology and quantum dimensions~\cite{DongEtAl2008,WenMatsuuraRyu2016Edge,BondersonKnappPatel2017,LoChang2024TorusKnots}. When a subsystem comprises multiple parallel annuli, these components probe a single shared flux distribution. Consequently, the state-dependent term need not reduce to a sum of independent boundary contributions. The torus therefore provides a unified topological framework in which the R\'enyi order, the spatial configuration, and the anyon data can be varied independently.

For finite positive orders other than one, no nontrivial homogeneous inequality between ordinary R\'enyi entropies is valid for all multipartite quantum states~\cite{LindenMosonyiWinter2013RenyiInequalities}. This no-go result leaves open spatial inequalities of the canonical torus functional within a fixed topological ground-state subspace. At the von Neumann point, Lo and Chang announced the strong subadditivity property for the torus topological entropy~\cite{LoChang2024TorusKnots} and reduced arbitrary subsurface pairs to ribbon configurations~\cite{LoChang2025TorusSSA}. Their single-crossing ribbon branch rests on a modular entropic inequality, tested numerically across available unitary categories up to rank eleven. The fusion-Fourier uncertainty relation of Liu, Palcoux, and Wu~\cite{LiuPalcouxWu2021Fusion} supplies the corresponding analytic bound. Related connections between spatial entanglement and anyonic data appear in studies of mutually dual flux bases~\cite{JianKimQi2015LRMI}, information-convex sets~\cite{ShiKatoKim2020}, and entanglement-bootstrap remote detectability~\cite{ShiHuangMcGreevy2025TorusTrick}. These developments motivate the classification question: how does universal strong subadditivity of the canonical torus functional depend on the entropy order and anyon theory when all pure logical states and admissible region pairs are included?

We resolve this question by classifying universal strong subadditivity of the canonical torus functional across all pure logical states and admissible region pairs. At every fixed $0<\alpha<1$, universal validity is equivalent to pointed, or Abelian, anyon structure. At $\alpha=1$, the inequality holds for every finite unitary modular tensor category through homological constraints and the existing weighted uncertainty bound. Unequal quantum dimensions curve the shared-flux contribution and give violations in every nonpointed category at every other finite positive order. Every category of rank greater than one also has a single state and region pair that violate the inequality throughout finite $\alpha\ge2$. The remaining pointed range $1<\alpha<2$ is governed by an exact equal-order modular Fourier criterion. We realize the nonpointed branches in ordinary physical-edge entropies of a specified quantum-double reference ground state for every finite non-Abelian group, using fixed region pairs on a common refined lattice. Applying the same bounded real diagonal filter to every physical edge then gives a controlled deformation. For each fixed group, one common small parameter interval supports a uniform R\'enyi-two estimate and an independently justified gapped local parent on sufficiently large bounded-aspect-ratio tori. Nonconstant filter functions give a non-isospectral family near zero at each fixed finite volume under the stated graph conditions. Its R\'enyi-two deficit converges to the strictly negative fixed-point value in the controlled joint limit of large system size and interface separation.

The remainder of the paper is organized as follows. In Sec.~\ref{sec:classification}, we define the canonical functional and state the entropy-order classification. We then analyze the shared-flux mechanism and the high-order pointed boundary in Sec.~\ref{sec:mechanism}, before proving the positive regimes across all admissible subsurfaces in Sec.~\ref{sec:geometry}. Turning to microscopic models, Secs.~\ref{sec:microscopic} and~\ref{sec:filtered-witness} construct the ordinary-edge realization and its controlled local deformation. Finally, Sec.~\ref{sec:discussion} examines the resulting constraints on topological entanglement, and Appendix~\ref{app:uncertainty} records the weighted Fourier normalization.

%% file: sections/objects.tex
\section{Canonical torus entropy and its classification}\label{sec:classification}

In the topological description, local boundary terms are separated from the nonlocal contribution of noncontractible cuts. The resulting canonical functional depends on the logical state and the anyon theory. We first specify its state and region domains, then state the entropy-order classification. Section~\ref{sec:microscopic} realizes this functional within the ordinary edge entropies of specified microscopic states.

\subsection{State space, regions and entropy prescription}
Our setting is the topological quantum field theory of a finite unitary modular tensor category $\mathcal C$ on an unpunctured spatial two-torus, where the canonical entropy functional records both the boundary topology of a cut and the logical flux it intercepts. Each simple anyon $a$ carries a quantum dimension $d_a\geq1$, with $d_0=1$ for the vacuum and total quantum dimension $D=\sqrt{\sum_a d_a^2}$; the category $\mathcal C$ is pointed, or Abelian, precisely when $d_a=1$ for every simple label. Selecting a primitive noncontractible cycle $\gamma$ specifies an orthonormal definite-flux basis $\{|a_\gamma\rangle\}$ of normalized Reshetikhin--Turaev states, each prepared on a solid torus with meridian $\gamma$ and core label $a$~\cite{Witten1989Jones,ReshetikhinTuraev1991Invariants}. In this basis, any pure ground state $\psi$ determines a sector probability distribution $p_a^\gamma=|\langle a_\gamma|\psi\rangle|^2$.

Spatial subsystems are modeled as compact regular subsurfaces of the torus, which may be disconnected and may contain contractible holes. An embedded subsurface $R$ is admissible if all of its noncontractible boundary circles lie parallel to a common primitive homology cycle $\gamma_R$, and we write $b_R$ for its total number of boundary components and $2m_R$ for its number of noncontractible boundary circles. When considering a pair of regions $A$ and $B$, we require their boundaries to be disjoint or to intersect transversely, rounding any corners so that both the union $A\cup B$ and the intersection $A\cap B$ remain regular subsurfaces. For such regions, we use the canonical topological R\'enyi functional following the surgery, edge-theory, and replica prescriptions~\cite{DongEtAl2008,WenMatsuuraRyu2016Edge,BondersonKnappPatel2017,LoChang2024TorusKnots,LoChang2025TorusSSA}
\begin{equation}
 S_\alpha^{\rm top}(R;\psi)=-b_R\log D+
 \begin{cases}F_\alpha(m_R;p^{\gamma_R}),&m_R>0,\\0,&m_R=0,\end{cases}
 \label{eq:canonical}
\end{equation}
where for finite $\alpha>0$ with $\alpha\ne1$,
\begin{equation}
 F_\alpha(m;p)=\frac{1}{1-\alpha}\log\sum_{a:p_a>0}p_a^\alpha d_a^{2m(1-\alpha)}.
 \label{eq:annular}
\end{equation}
Equation~\eqref{eq:annular} defines the real-order continuation of $F_\alpha$ via a finite sum over the positive support of $p$, while its von Neumann value at $\alpha=1$ is obtained from the limit $\alpha\to1$ using natural logarithms and the convention $0\log0=0$. Whenever $m_R=0$, the subsurface $R$ has no noncontractible boundary circles and the flux-dependent term in Eq.~\eqref{eq:canonical} drops out even if $R$ itself wraps the torus. For a contractible disk, $S_\alpha^{\rm top}$ therefore reduces to the $\alpha$-independent topological entanglement entropy~\cite{FlammiaHammaHughesWen2009TopologicalRenyi}, whereas cuts with $m_R>0$ remain sensitive to the global topological flux threading the torus~\cite{ZhangEtAl2012MES}. Because $S_\alpha^{\rm top}$ is not in general the R\'enyi entropy of a standalone reduced density matrix, its strong-subadditivity properties must be established directly from Eqs.~\eqref{eq:canonical} and~\eqref{eq:annular}.

\subsection{Universal strong subadditivity}
For any pure logical state $\psi$ and admissible region pair $(A,B)$, the order-$\alpha$ strong-subadditivity deficit of the canonical functional is
\begin{equation}
 \Delta_\alpha=S_\alpha^{\rm top}(A)+S_\alpha^{\rm top}(B)
 -S_\alpha^{\rm top}(A\cup B)-S_\alpha^{\rm top}(A\cap B).
 \label{eq:deficit}
\end{equation}
Universal strong subadditivity holds at order $\alpha$ when $\Delta_\alpha\ge 0$ across every pure logical state $\psi$ and every admissible pair $(A,B)$. This joint quantification over all states and cuts is indispensable because an individual state can yield a vanishing deficit even in a nonpointed category $\mathcal C$.

\begin{theorem}[Entropy order and anyon structure]\label{thm:classification}
For the canonical functional of a finite unitary modular tensor category on the unpunctured torus, the following statements hold.
\begin{enumerate}
\item At every fixed order strictly between zero and one, universal strong subadditivity holds if and only if the category is pointed.
\item At the von Neumann order, universal strong subadditivity holds for every such category.
\item Every nonpointed category admits a violation at each finite positive order other than one. In a fixed pointed category, universal validity above order one is equivalent to the entropic Fourier bound below for every normalized logical state.
\end{enumerate}
\end{theorem}
\begin{equation}
\begin{gathered}
H_\alpha(|\psi|^2)+H_\alpha(|S\psi|^2)\geq\log r
 \\\text{for every normalized }\psi.
\end{gathered}
\label{eq:pointed-high}
\end{equation}

In Eq.~\eqref{eq:pointed-high}, $r=D^2$ is the rank of the pointed category and $H_\alpha$ denotes the standard order-$\alpha$ R\'enyi entropy of a probability distribution. For two primitive cycles with intersection number one, the modular matrix $S$ connects the corresponding matched flux bases, sending the normalized amplitude vector $\psi$ in one basis to $S\psi$ in the dual basis. Unlike the $\alpha<1$ criterion, which separates Abelian from non-Abelian theories, the condition for $\alpha>1$ probes the modular Fourier transform within a given pointed category. Section~\ref{sec:mechanism} constructs the violating geometries and explains their physical mechanism, while Section~\ref{sec:geometry} establishes the positive regimes across all admissible subsurfaces.

\begin{corollary}[Universal high-order failure]\label{cor:high-orders}
For every finite unitary modular tensor category of rank greater than one, there is a pure logical state and an admissible region pair with $\Delta_\alpha<0$ at every finite order $\alpha\ge2$.
\end{corollary}
A single state and region pair suffice throughout $\alpha\ge2$. Section~\ref{sec:mechanism} proves the pointed case with the self-dual witness in Eq.~\eqref{eq:pointed-witness} and the nonpointed case with the shared-flux witnesses. The remaining fixed-category question is which orders in $1<\alpha<2$ satisfy the Fourier criterion.

This classification addresses a constrained topological problem rather than the unconstrained multipartite setting of Linden, Mosonyi, and Winter~\cite{LindenMosonyiWinter2013RenyiInequalities}. Whereas general multipartite R\'enyi entropies satisfy no nontrivial homogeneous inequalities away from $\alpha=1$, Theorem~\ref{thm:classification} restricts states to a fixed topological logical space and evaluates the geometric functional~\eqref{eq:canonical}, whose four-term linear combination~\eqref{eq:deficit} also differs from divergence-based R\'enyi conditional mutual informations~\cite{BertaSeshadreesanWilde2015RenyiCMI}.

Lo and Chang announced the torus topological-entropy SSA property in Ref.~\cite{LoChang2024TorusKnots}. Their later analysis reduces general region pairs to ribbons and isolates a modular uncertainty condition on the once-crossing branch~\cite{LoChang2025TorusSSA}. The fusion-Fourier theorem of Liu, Palcoux, and Wu~\cite{LiuPalcouxWu2021Fusion} supplies the required analytic bound. Section~\ref{sec:geometry} combines it with an independent homological analysis of the actual union and intersection. This gives the global geometric input for the R\'enyi classification.

%% file: sections/mechanism.tex
\section{Shared flux and the mechanism of R\'enyi violations}\label{sec:mechanism}

When a category is nonpointed, the violations in Theorem~\ref{thm:classification} arise because multiple boundary circles couple to the same global flux distribution. Parallel annuli do not sample independent random variables; instead, they probe a single shared topological charge. If the occupied sectors have unequal quantum dimensions, the resulting dimensional weights make $F_\alpha(m;p)$ nonlinear in the annulus count $m$. We construct two explicit strip configurations on the torus to isolate the two opposite signs of its second difference.

\subsection{Quantum-dimension curvature}
In the von Neumann limit $\alpha\to1$, the annular functional in Eq.~\eqref{eq:annular} becomes
\begin{equation}
\begin{gathered}
F_1(m;p)=H(p)+2m\sum_a p_a\log d_a,
 \\ H(p)=-\sum_a p_a\log p_a.
\end{gathered}
\label{eq:order-one}
\end{equation}
Here $F_1(m;p)$ separates into the Shannon entropy $H(p)$ and a quantum-dimension average that grows linearly with $m$ at slope $2\sum_a p_a\log d_a$. Away from $\alpha=1$, the factor $d_a^{2m(1-\alpha)}$ in Eq.~\eqref{eq:annular} alters the relative weights of sectors with unequal quantum dimensions as $m$ changes. Even for a fixed physical state $\psi$, varying the annulus count reweights the sector sum. To compute the derivatives of $F_\alpha(m;p)$ at a fixed distribution $p$, we set $x_a=\log d_a$ on the occupied support and treat $m$ as a continuous real variable:
\begin{equation}
\begin{gathered}
w_a^{(m)}=\frac{p_a^\alpha e^{2m(1-\alpha)x_a}}{\sum_c p_c^\alpha e^{2m(1-\alpha)x_c}},\\
 \partial_m^2F_\alpha=4(1-\alpha)\operatorname{Var}_{w^{(m)}}(x).
\end{gathered}
\label{eq:curvature}
\end{equation}
Whenever the occupied support contains more than one value of $d_a$, the variance $\operatorname{Var}_{w^{(m)}}(x)$ is strictly positive for all $m$. Consequently, $F_\alpha(m;p)$ is strictly convex in $m$ for $0<\alpha<1$ and strictly concave for $\alpha>1$. If all occupied sectors share the same quantum dimension, the variance vanishes and $F_\alpha(m;p)$ remains affine in $m$.

\subsection{Two spatial witnesses}
Once a logical state and a common cut basis occupy sectors with unequal quantum dimensions, this curvature directly produces negative strong-subadditivity deficits. We parametrize the torus as $[0,12)\times S^1$ with periodic boundary conditions on $[0,12)$ and define two pairs of longitudinal strips (Fig.~\ref{fig:annulus-geometries}). For $0<\alpha<1$, we choose
\begin{equation}
 A_-=([0,2]\cup[3,5])\times S^1,\qquad
 B_-=([1,4]\cup[\tfrac92,6])\times S^1.
 \label{eq:regions-low}
\end{equation}
The boundary circles of $A_-$ and $B_-$ are mutually disjoint. Across the quadruple $(A_-,\allowbreak B_-;\allowbreak A_-\cup B_-,\allowbreak A_-\cap B_-)$, the annulus counts are $(m_{A_-},\allowbreak m_{B_-};\allowbreak m_{A_-\cup B_-},\allowbreak m_{A_-\cap B_-})=(2,2;1,3)$, whereas the total boundary counts satisfy $b_{A_-}+b_{B_-}=b_{A_-\cup B_-}+b_{A_-\cap B_-}=8$. Writing $\Delta b=b_A+b_B-b_{A\cup B}-b_{A\cap B}$ for the net boundary-component difference, we have $\Delta b=0$. The $-b_R\log D$ contributions in Eq.~\eqref{eq:deficit} therefore cancel, leaving a pure second difference in $m$:
\begin{equation}
\begin{gathered}
\Delta_\alpha(A_-,B_-)=2F_\alpha(2;p)-F_\alpha(1;p)-F_\alpha(3;p)<0
 \\(0<\alpha<1).
\end{gathered}
\label{eq:negative-low}
\end{equation}
For $\alpha>1$, we invert the sign of this second difference with a second strip pair,
\begin{equation}
 A_+=[1,5]\times S^1,\qquad
 B_+=([0,2]\cup[4,6]\cup[8,9])\times S^1.
 \label{eq:regions-high}
\end{equation}
These subsurfaces also have disjoint boundaries with $\Delta b=0$, but their annulus counts are $(1,3;2,2)$. By the strict concavity of $F_\alpha(m;p)$ for $\alpha>1$, the deficit is again negative:
\begin{equation}
\begin{gathered}
\Delta_\alpha(A_+,B_+)=F_\alpha(1;p)+F_\alpha(3;p)-2F_\alpha(2;p)<0
 \\(\alpha>1).
\end{gathered}
\label{eq:negative-high}
\end{equation}
Because every nonpointed category has at least one simple sector $b$ with $d_b>1$, the equal superposition $\psi=(|0_\gamma\rangle+|b_\gamma\rangle)/\sqrt2$ occupies two distinct quantum dimensions. This state yields $\Delta_\alpha(A_-,B_-)<0$ for $0<\alpha<1$ from Eq.~\eqref{eq:negative-low} and $\Delta_\alpha(A_+,B_+)<0$ for $\alpha>1$ from Eq.~\eqref{eq:negative-high}. The modular-transformed vacuum $S|0\rangle$, which has full support with $p_a=d_a^2/D^2$, provides a second witness that connects directly to the microscopic states in Sec.~\ref{sec:microscopic}. Conversely, if a state $\psi$ is supported entirely on sectors of equal quantum dimension, then $\operatorname{Var}_{w^{(m)}}(x)=0$ and $F_\alpha(m;p)$ is affine in $m$. Both second differences then vanish for that state, even though the category itself violates universal strong subadditivity.

Expanding around $\alpha=1$ quantifies how these violations emerge for a fixed distribution $p$ on its finite occupied support. With $t=\alpha-1$ and $\ell_p=\sum_a p_a\log d_a$, the finite sector sum expands as
\begin{equation}
 F_{1+t}(m;p)=H(p)+2m\ell_p
 -\frac t2\operatorname{Var}_p(\log p_a-2m\log d_a)+O(t^2).
 \label{eq:near-one-functional}
\end{equation}
Forming the spatial second differences eliminates both $\operatorname{Var}_p(\log p_a)$ and the covariance between $\log p_a$ and $\log d_a$:
\begin{equation}
\begin{gathered}
 \Delta_-(1+t)=4t\operatorname{Var}_p(\log d_a)+O(t^2),\\
 \Delta_+(1+t)=-4t\operatorname{Var}_p(\log d_a)+O(t^2).
\end{gathered}
\label{eq:near-one-deficits}
\end{equation}
Here $\Delta_\mp$ denote the deficits of the two strip pairs $(A_\mp,B_\mp)$. Selecting the negative branch on either side of $\alpha=1$ produces a leading deficit of $-4|\alpha-1|\operatorname{Var}_p(\log d_a)$, showing that fluctuations in the occupied quantum dimensions govern the initial linear response in $\alpha-1$.

Changing the common primitive cut direction need not eliminate the equal-dimension blind spot, as a compact Ising example illustrates. In the standard basis $(\mathbf1,\varepsilon,\sigma)$ with quantum dimensions $(1,1,\sqrt2)$, let $u_\pm=(|\mathbf1\rangle\pm|\varepsilon\rangle)/\sqrt2$ and $u_\sigma=|\sigma\rangle$. Up to an overall framing phase, the standard modular transformations~\cite{Kitaev2006} act as
\begin{equation}
\begin{gathered}
 Su_+=u_+,\quad Su_-=u_\sigma,\quad Su_\sigma=u_-,\\
 Tu_+=u_-,\quad Tu_-=u_+,\quad Tu_\sigma=e^{i\pi/8}u_\sigma.
\end{gathered}
\label{eq:ising-orbit}
\end{equation}
The modular generators and their inverses permute these three rays, and each ray is supported on sectors of a single quantum dimension. Because modular basis changes relate all primitive cut directions on the torus, both parallel-annulus second differences vanish for each of $u_+$, $u_-$, and $u_\sigma$ along every primitive cut direction and at every finite order $\alpha>0$. The example shows the blind spot of these two parallel-annulus witness families, consistently with the all-state quantifier of Theorem~\ref{thm:classification}.

The condition $\Delta b=0$ removes the canonical boundary-count term in both strip pairs. In Section~\ref{sec:microscopic}, we show that the nonuniversal boundary terms in full microscopic entropies likewise cancel for specified states and cuts on a common lattice refinement. Before turning to those microscopic realizations, we analyze how intersecting cuts in pointed categories violate strong subadditivity for $\alpha>1$.

\input{sections/figure.tex}

\subsection{High orders in pointed categories}

When two single annuli cross once in a pointed category, the deficit is the sum of the R\'enyi entropies in the two cycle bases minus the logarithm of the category's rank. The analytic setting is that of complementary-basis and noncommutative R\'enyi uncertainty relations~\cite{BialynickiBirula2006RenyiUncertainty,LiuPalcouxWu2021Fusion}. Earlier studies analyzed equal-order uncertainty bounds and optimizing states through majorization~\cite{PuchalaRudnickiZyczkowski2013Majorization} and finite-dimensional optimization~\cite{AbdelkhalekEtAl2015Optimality}, including explicit numerical and semi-analytical results for the qubit Hadamard transform~\cite{ZozorBosykPortesi2013QubitRenyi}. For the torus classification, we do not need the optimal threshold in $\alpha$; simple analytical witness states suffice.

Above order one, the modular Fourier inequality~\eqref{eq:pointed-high} can fail. In a pointed category of rank $N$, the vacuum column of $S$ has uniform entries $S_{a0}=N^{-1/2}$ and satisfies $S^2e_0=e_0$. The normalized superposition
\begin{equation}
 \psi=\frac{e_0+Se_0}{\sqrt{2+2/\sqrt N}},\qquad S\psi=\psi
 \label{eq:pointed-witness}
\end{equation}
is a $+1$ eigenstate of $S$, so the two cycle bases have the same probability distribution,
\begin{equation}
 p_0=\frac{1+N^{-1/2}}2,\qquad
 p_{a\ne0}=\frac1{2N(1+N^{-1/2})}.
 \label{eq:pointed-witness-probabilities}
\end{equation}
Because $p_0>1/2$, the R\'enyi entropy at any $\alpha>1$ is bounded above by
\begin{equation}
 H_\alpha(p)<\frac{\alpha}{\alpha-1}\log2.
 \label{eq:pointed-entropy-upper}
\end{equation}
The sum $H_\alpha(p)+H_\alpha(q)=2H_\alpha(p)$ therefore drops strictly below $\log N$ whenever
\begin{equation}
 \log N>\frac{2\alpha}{\alpha-1}\log2.
 \label{eq:pointed-rank-condition}
\end{equation}
Since Deligne tensor powers of any nontrivial pointed category achieve arbitrarily large rank, Eq.~\eqref{eq:pointed-rank-condition} provides a pointed counterexample at every fixed $\alpha>1$, including the interval $1<\alpha<2$.

At $\alpha=2$, this witness rules out universal strong subadditivity for every nontrivial pointed category without increasing the rank. Its collision probability is
\begin{equation}
 C_N=\sum_a p_a^2
 =\frac{N+3\sqrt N+4}{4\sqrt N(\sqrt N+1)}.
 \label{eq:pointed-collision}
\end{equation}
For a pair of once-crossing annuli, the order-two deficit evaluates to
\begin{equation}
 \Delta_2=2H_2(p)-\log N
 =2\log\frac{4(\sqrt N+1)}{N+3\sqrt N+4}<0\qquad(N>1).
 \label{eq:pointed-all-ranks}
\end{equation}
The strict inequality holds because the denominator exceeds the numerator by $N-\sqrt N>0$. By the monotonicity of $H_\alpha(p)$ in $\alpha$, the same state and region pair satisfy $\Delta_\alpha\le\Delta_2<0$ for every finite $\alpha\ge2$. Combining this pointed witness with the nonpointed shared-flux witnesses establishes Corollary~\ref{cor:high-orders}. Specializing Eq.~\eqref{eq:pointed-all-ranks} to $N=2$ and $N=4$ gives $\Delta_2=\log(8/9)$ for the semion theory and $\Delta_2=\log(36/49)$ for the toric code, whereas $N=1$ yields a vanishing canonical functional. The collision bound settles universal failure from order two upwards. Deciding which orders in $1<\alpha<2$ satisfy the Fourier criterion for a fixed pointed category remains a separate optimization problem.

%% file: sections/figure.tex
\begin{figure*}[t]
 \centering
 \includegraphics[width=\linewidth]{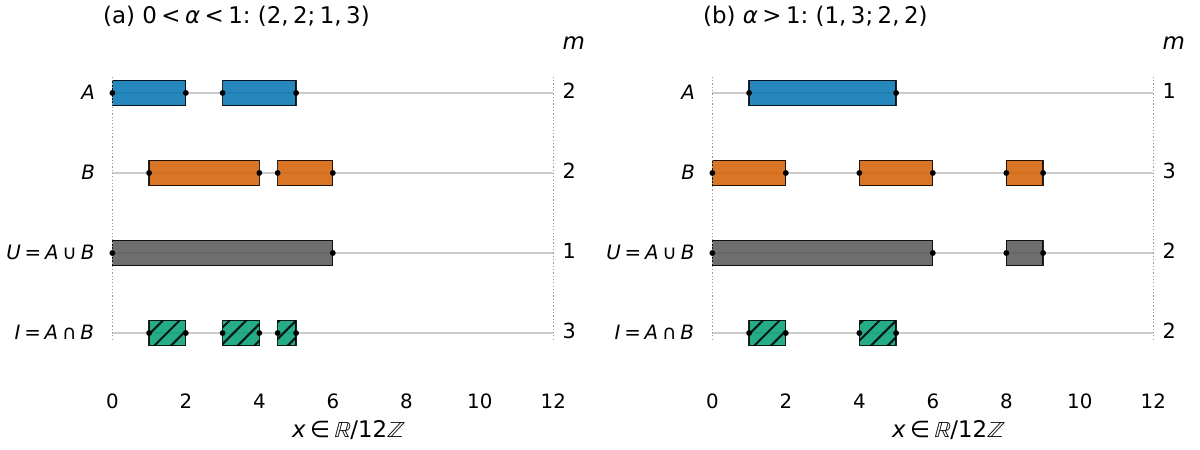}
\caption{Parallel-annulus configurations on the torus. Each strip represents the product of a closed transverse interval in $\mathbb R/12\mathbb Z$ with the longitudinal circle $S^1$. Rows show the input regions $A,B$, their union $U$, and their intersection $I$, with the respective annulus counts $m_R$ indicated on the right. The boundary circles of $A$ and $B$ are pairwise disjoint, although the regions overlap. Configuration~(a) has counts $(2,2;1,3)$ and isolates minus the second difference of $F_\alpha$, as in equation~\eqref{eq:negative-low}. Configuration~(b) has counts $(1,3;2,2)$ and isolates plus the same second difference, as in equation~\eqref{eq:negative-high}. The boundary terms cancel in both cases. When occupied sectors have different quantum dimensions, both deficits are strictly negative in their indicated order ranges.}
 \label{fig:annulus-geometries}
\end{figure*}

%% file: sections/positive.tex
\section{Geometry, uncertainty and the positive regimes}\label{sec:geometry}

Establishing the positive regimes of Theorem~\ref{thm:classification} requires tracking how the actual union $U=A\cup B$ and intersection $I=A\cap B$ link boundary counts to logical sector entropies across all admissible subsurfaces. In the sole geometric branch where the boundary contribution to the deficit can be negative, integral homology forces the two inputs to carry essential annuli along complementary primitive slopes. A weighted entropic uncertainty relation between those dual flux bases then supplies the exact positive contribution needed to offset the negative topological term. The subsections below develop this complete surface classification, including disconnected regions and contractible holes.

This mechanism differs from related strong-subadditivity arguments on the torus and on the plane. Earlier torus analyses use logical mixtures~\cite{Kim2013} or flux-basis uncertainty~\cite{JianKimQi2015LRMI}. The cited planar results assume exact area laws~\cite{BaoCaoWalter2015EtAl} or nondegenerate ground states~\cite{NaskarSamal2026}. Exact-area-law and zero-correlation-length assumptions also link topological entanglement entropy to irreducible correlations~\cite{KatoFurrerMurao2016Information}. Area laws additionally constrain local boundary descriptions~\cite{KatoBrandao2019EdgeLocality}. Because the canonical torus functional is defined directly at the level of the topological model rather than as an ordinary reduced-state entropy of an arbitrary microscopic system, its spatial inequality requires an independent homological proof. Likewise, our linear four-region R\'enyi deficit is distinct from divergence-based R\'enyi conditional mutual information~\cite{BertaSeshadreesanWilde2015RenyiCMI}, so positivity results for the latter do not govern the present classification.

\subsection{Surface types and component counting}
Each connected subsurface of the torus has an ambient homology image in $H_1(T^2;\mathbb Z)$ of rank zero, one or two, which we denote by types $P$, $E$ and $G$, respectively. For any planar component, this image is generated by its boundary circles and is therefore isotropic in $H_1(T^2;\mathbb Z)$. When the image is nonzero, it is generated by a single primitive class. Because disjoint essential circles on $T^2$ are necessarily parallel, cutting the torus along them yields annular strips. Every connected planar component with a nonzero ambient homology image thus possesses two essential boundary circles, together with any number of contractible holes; these are the components of type $E$. Planar components whose ambient homology image vanishes have type $P$.

By contrast, a component of genus one contains a pair of curves with intersection number one and therefore spans the entire ambient lattice $H_1(T^2;\mathbb Z)$. No disjoint essential circle can avoid both curves, which implies that an admissible region $R$ contains at most one $G$ component and can never combine a $G$ component with an $E$ component. Moreover, all $E$ components within $R$ must share a common primitive slope. If $c_R$ denotes the total number of connected components of $R$ and $g_R\in\{0,1\}$ its genus, the counts of $E$, $G$ and $P$ components are given by $m_R$, $g_R$ and $c_R-m_R-g_R$, respectively.

For any regional quantity $f$, write $\Delta f=f_A+f_B-f_U-f_I$. We form the bipartite incidence graph whose vertices are the connected components of $A$ and $B$ and whose edges are the connected components of $I$, each edge connecting the input components that contain it. Because the connected components of this graph correspond bijectively to the connected components of $U$, its cycle rank is
\begin{equation}
 \beta=c_I-c_A-c_B+c_U=-\Delta c\ge0.
 \label{eq:incidence-beta}
\end{equation}
Combining Euler characteristic additivity with the surface relation $\chi(R)=2c_R-2g_R-b_R$ yields the boundary-component deficit
\begin{equation}
 \Delta b=-2X,\qquad X=\beta+\Delta g\ge-1.
 \label{eq:boundary-X}
\end{equation}
Here the genus deficit satisfies $\Delta g\ge-1$. If $g_I=1$, inclusion forces all four regions to have genus one and $\Delta g=0$. If instead $g_I=0$, only $g_U=1$ can make $\Delta g$ negative, and then at most $-1$. The integer invariant $X$ can therefore be negative only when $X=-1$.

\subsection{A local-system constraint and the negative branch}
To relate the annulus counts $m_R$ to $X$, choose a unitary character $\eta:\mathbb Z^2\to U(1)$ that is nontrivial on every nonzero vector of the ambient lattice,
\begin{equation}
 \eta(u,v)=\exp\!\bigl(2\pi i(\theta_1u+\theta_2v)\bigr),
 \qquad 1,\theta_1,\theta_2\ \text{rationally independent}.
 \label{eq:generic-character}
\end{equation}
Let $\mathcal L_\eta$ denote the associated rank-one local system on $T^2$, introduced purely as a homological book-keeping device rather than a physical gauge flux. On any connected component of $R$, the zeroth homology with coefficients in $\mathcal L_\eta$ is one-dimensional if and only if the restricted holonomy is trivial. This holds on components of type $P$ and fails on those of type $E$ or $G$, so
\begin{equation}
 \dim_{\mathbb C}H_0(R;\mathcal L_\eta)=c_R-m_R-g_R.
 \label{eq:twisted-H0}
\end{equation}
Admissibility allows us to pass to a common cellular refinement or to compatible collars that deformation retract onto $A,B,U,I$. The terminal segment of the Mayer--Vietoris sequence with local coefficients reads
\begin{equation}
 H_0(I;\mathcal L_\eta)\longrightarrow
 H_0(A;\mathcal L_\eta)\oplus H_0(B;\mathcal L_\eta)
 \longrightarrow H_0(U;\mathcal L_\eta)\longrightarrow0.
 \label{eq:twisted-MV}
\end{equation}
Exactness of Eq.~\eqref{eq:twisted-MV} implies $\Delta(c-m-g)\le0$, which rewrites directly as
\begin{equation}
 X+\Delta m\ge0.
 \label{eq:annulus-constraint}
\end{equation}

When $X=-1$, Eq.~\eqref{eq:boundary-X} forces $\beta=0$, $g_A=g_B=g_I=0$ and $g_U=1$. Because $\beta=0$, the component incidence graph is a forest and its signed incidence map $H_0(I;\mathbb Z)\to H_0(A;\mathbb Z)\oplus H_0(B;\mathbb Z)$ is injective. The integral Mayer--Vietoris sequence, together with the presence of a genus-one component in $U$, then yields the composite surjection
\begin{equation}
 H_1(A;\mathbb Z)\oplus H_1(B;\mathbb Z)
 \twoheadrightarrow H_1(U;\mathbb Z)
 \twoheadrightarrow H_1(T^2;\mathbb Z).
 \label{eq:integral-surjection}
\end{equation}
Since $g_A=g_B=0$, each input region is planar and its ambient homology image is either trivial or generated by a single primitive slope. Generating the full integral lattice $H_1(T^2;\mathbb Z)$ via Eq.~\eqref{eq:integral-surjection} therefore requires both inputs to carry essential annuli whose primitive slopes $\gamma_A$ and $\gamma_B$ form an integral basis:
\begin{equation}
 \det(\gamma_A,\gamma_B)=\pm1,\qquad
 m_A,m_B\ge1,\qquad m_U=m_I=0.
 \label{eq:negative-branch}
\end{equation}
Here $m_U=0$ because $g_U=1$ precludes $E$ components in the union, while $m_I=0$ because any essential cycle in the intersection would have to belong simultaneously to the two distinct rank-one slope subgroups of $A$ and $B$. Integral surjectivity is essential here, as a merely nonzero intersection number would not force $\det(\gamma_A,\gamma_B)=\pm1$.

We now orient the two primitive slopes so that their intersection number is $+1$ and map them by an orientation-preserving torus homeomorphism to the standard meridian and longitude. Under the Reshetikhin--Turaev normalization, the Hopf link invariant in $S^3$ is $Z(S^3;\operatorname{Hopf}(a,b))=S_{ab}$ with $Z(S^3)=1/D$~\cite{Witten1989Jones,ReshetikhinTuraev1991Invariants}. If the modular transformation between the two slope bases is modified by diagonal phase matrices to $D_1SD_2$, the right factor $D_2$ can be absorbed into the input state coordinates, whereas the left factor $D_1$ drops out of the output sector probabilities. Reversing slope orientations or applying charge conjugation likewise preserves the quantum-dimension weights. This fixes the matched $S$-related probability distributions $p_a=|\psi_a|^2$ and $q_a=|(S\psi)_a|^2$ for a given logical state without claiming that $q$ is invariant under rephasing the components of a fixed coordinate vector $\psi$.

\subsection{The sector-entropy bounds}
To control the annular contributions on the two dual slopes, we associate to any sector probability distribution $p$ its Shannon entropy $h_p=H(p)$ and its expected logarithmic quantum dimension $\ell_p=\sum_a p_a\log d_a$. With $L=\log D$ and $d_a\ge 1$, these quantities obey the elementary inequalities
\begin{equation}
 0\le\ell_p\le L,\qquad h_p\ge0,\qquad h_p+2\ell_p\le2L.
 \label{eq:sector-bounds}
\end{equation}
The ceiling $h_p+2\ell_p\le 2L$ reflects the nonnegativity of the Kullback--Leibler divergence of $p$ relative to the quantum-dimension distribution $r_a=d_a^2/D^2$,
\begin{equation}
 D_{\mathrm{KL}}(p\Vert r)=2L-h_p-2\ell_p\ge0.
 \label{eq:sector-KL}
\end{equation}
This single-distribution bound depends only on Eq.~\eqref{eq:sector-KL}, even though the same combination $h_p+2\ell_p$ also governs sector-level terms in information-convexity studies of topological entanglement~\cite{Shi2019InformationConvex,ShiKatoKim2020}.

When a normalized logical state vector $\psi$ is resolved in matched flux bases for two primitive cycles that intersect once, the resulting sector distributions $p_a=|\psi_a|^2$ and $q_a=|(S\psi)_a|^2$ cannot both equal the vacuum distribution $\delta_0$ when $D>1$. Theorem~4.9 of Liu, Palcoux and Wu~\cite{LiuPalcouxWu2021Fusion} quantifies this obstruction through the dimension-weighted uncertainty relation
\begin{equation}
 H(p)+H(q)+2\ell_p+2\ell_q\ge2L.
 \label{eq:weighted-uncertainty}
\end{equation}
Equality holds when the state lies in the vacuum sector of the first basis, for which $p=\delta_0$ and $q=r$. We give a self-contained Riesz--Thorin proof and the explicit trace dictionary relating Eq.~\eqref{eq:weighted-uncertainty} to Fourier uncertainty on fusion bialgebras~\cite{JiangLiuWu2016Uncertainty,MaassenUffink1988} in Section~\ref{app:uncertainty}.

For basic transverse cuts on the torus, Zhang et al.~\cite{ZhangEtAl2012MES} derived the same quantitative bound by combining their minimally entangled state transformation with microscopic strong subadditivity and the cancellation of local boundary terms. Because our target is the canonical topological functional itself rather than a reduced-state entropy of a microscopic lattice model, we instead use Eq.~\eqref{eq:weighted-uncertainty} as a direct algebraic consequence of the modular $S$ matrix. The uncertainty step in the negative-$X$ branch therefore requires no prior assumption that the canonical functional obeys strong subadditivity.

\subsection{Proof of the order-one inequality}
When $D=1$ the category is trivial and every term in $\Delta_1$ vanishes identically. For $D>1$, we establish $\Delta_1\ge0$ across all geometric branches without dividing by $L$.

\paragraph{The branch $X=-1$.}
By Eq.~\eqref{eq:negative-branch}, neither $U$ nor $I$ carries an $E$ component, so the output sector terms drop out. Since $m_A,m_B\ge1$ and the expectations $\ell_p,\ell_q$ are nonnegative, Eq.~\eqref{eq:weighted-uncertainty} gives
\begin{equation}
 \begin{split}
 \Delta_1
 &=-2L+H(p)+H(q)+2m_A\ell_p+2m_B\ell_q\\
 &\ge-2L+H(p)+H(q)+2\ell_p+2\ell_q\ge0.
 \end{split}
 \label{eq:proof-negative}
\end{equation}
A single essential annulus in each input already supplies the dimension weights required by the uncertainty relation, and any additional parallel annuli contribute further nonnegative multiples of $\ell_p$ and $\ell_q$.

\paragraph{Different input slopes with $X\ge0$.}
Suppose $X\ge0$ and the two inputs contain $E$ components of distinct primitive slopes $\gamma_A\ne\pm\gamma_B$. Their union $U$ then has a rank-two ambient homology image and must contain a $G$ component, whereas their intersection $I$ cannot contain an essential cycle lying in both slope subgroups. Consequently $m_U=m_I=0$, and the deficit reduces to $\Delta_1=2XL+2m_A\ell_{\gamma_A}+H(p^{\gamma_A})+2m_B\ell_{\gamma_B}+H(p^{\gamma_B})\ge0$. Every piece is individually nonnegative, so no uncertainty relation is needed even when the two slopes intersect multiple times.

\paragraph{A common slope with $X\ge0$.}
In all remaining configurations with $X\ge0$, inclusion of ambient homology images ensures that any $E$ component present in an input fixes the slope of every $E$ component in $U$ and $I$. If neither input contains an $E$ component, two $P$-only inputs can generate $E$ components only in $U$, two inputs containing $G$ components can generate them only in $I$, and a $P$-only input paired with one containing $G$ generates them in neither. Hence all state-bearing regions (those with $m_R>0$) share a single primitive slope. When all four regions have $m_R=0$, the deficit is simply $2XL\ge0$. Otherwise, let $\ell$ and $h$ denote the sector parameters along the common slope and define
\begin{equation}
 M=\Delta m,\qquad
 \kappa=\mathbf1_{m_A>0}+\mathbf1_{m_B>0}
 -\mathbf1_{m_U>0}-\mathbf1_{m_I>0}.
 \label{eq:common-counts}
\end{equation}
The order-one deficit then takes the form
\begin{equation}
 \Delta_1=2XL+2M\ell+\kappa h.
 \label{eq:common-deficit}
\end{equation}
When $\kappa\ge0$ and $M\ge0$, Eq.~\eqref{eq:common-deficit} is manifestly nonnegative term by term. When $\kappa\ge0$ and $M<0$, applying the homological constraint $M\ge-X$ from Eq.~\eqref{eq:annulus-constraint} gives
\begin{equation}
 \Delta_1\ge2X(L-\ell)+\kappa h\ge0
 \qquad(\kappa\ge0).
 \label{eq:common-positive-kappa}
\end{equation}
The final step uses $\ell\le L$ from Eq.~\eqref{eq:sector-bounds}.

It remains to examine $\kappa<0$. If both $U$ and $I$ are state-bearing, then $U$ and hence both inputs are planar; the presence of an essential cycle in $I$ then forces both $A$ and $B$ to be state-bearing as well, which yields $\kappa=0$. Thus $\kappa$ can be negative only when at most one output is state-bearing, which requires $m_A=m_B=0$ and a single output with $m=k\ge1$. This fixes $\kappa=-1$ and $M=-k$, while Eq.~\eqref{eq:annulus-constraint} guarantees $X\ge k$. Decomposing $\Delta_1$ via the sector bounds in Eq.~\eqref{eq:sector-bounds} then gives
\begin{equation}
 \begin{split}
 \Delta_1&=2XL-2k\ell-h\\
 &=2(X-k)L+2(k-1)(L-\ell)+(2L-2\ell-h)\ge0.
 \end{split}
 \label{eq:common-negative-kappa}
\end{equation}
This completes the proof of $\Delta_1\ge0$ across all admissible torus subsurfaces. At every finite positive R\'enyi order $\alpha\ne1$, by contrast, suitable parallel-annulus profiles convert the strict cumulant curvature on states supported on unequal quantum dimensions into a negative deficit, as shown in Eqs.~\eqref{eq:curvature}--\eqref{eq:negative-high}.

\subsection{Geometric cases and vacuum saturation}\label{sec:geometry-guide}
Table~\ref{tab:geometry-guide} summarizes how the homological and entropic ingredients combine across the geometric branches. Although disconnected regions may adjoin $P$ components to either $E$ or $G$ components, a single region never combines $E$ and $G$; the integer counts $(X,m_A,m_B,m_U,m_I)$ are therefore constrained by surface topology rather than varying independently.

\begin{table}[htbp]
\caption{Geometric branches of the canonical order-one inequality. The common-slope rows apply after separating different input slopes and the case with no $E$ component in any of the four regions.}
\label{tab:geometry-guide}
\begingroup
\setlength{\tabcolsep}{5pt}
\begin{tabular}{p{0.20\linewidth}p{0.27\linewidth}p{0.45\linewidth}}
\hline
Branch & Geometric constraints & Non-negativity \\
\hline
$D=1$ & Trivial category & Every entropy contribution vanishes. \\
$X=-1$ & $\det(\gamma_{\textstyle A},\gamma_{\textstyle B})=\pm1$; $m_{\textstyle A},m_{\textstyle B}\ge1$, $m_{\textstyle U}=m_{\textstyle I}=0$ & Integral surjectivity, Eq.~\eqref{eq:negative-branch}, and weighted uncertainty, Eq.~\eqref{eq:proof-negative}. \\
$X\ge0$, different input slopes & $m_{\textstyle U}=m_{\textstyle I}=0$ & The boundary and input sector terms are individually non-negative. \\
$X\ge0$, no $E$ component & $m_{\textstyle A}=m_{\textstyle B}=m_{\textstyle U}=m_{\textstyle I}=0$ & Only the boundary term $2XL\ge0$ remains. \\
$X\ge0$, common slope; $\kappa\ge0$, $M\ge0$ & Common sector distribution & Each term in Eq.~\eqref{eq:common-deficit} is non-negative. \\
$X\ge0$, common slope; $\kappa\ge0$, $M<0$ & $X+M\ge0$ & Use $\ell\le L$ in Eq.~\eqref{eq:common-positive-kappa}. \\
$X\ge0$, common slope; $\kappa=-1$ & Neither input is state-bearing; one output has $m=k\ge1$, so $M=-k$ and $X\ge k$ & The sector bound $h+2\ell\le2L$ yields the three-term decomposition in Eq.~\eqref{eq:common-negative-kappa}. \\
\hline
\end{tabular}
\endgroup
\end{table}

In the basic configuration of two single annuli crossing once, one has $X=-1$, $m_A=m_B=1$ and $m_U=m_I=0$. Preparing the vacuum flux along $A$ yields $p=\delta_0$ and $q_a=d_a^2/D^2$, which saturates Eq.~\eqref{eq:weighted-uncertainty} and produces $\Delta_1=0$ under the trace convention of Eq.~\eqref{eq:LPW-trace-dictionary}.

\subsection{Pointed categories at other orders}
In any pointed category of rank $r=D^2$, every simple object has unit quantum dimension ($d_a=1$), so the sector functional $F_\alpha(m;p)$ reduces to the standard R\'enyi entropy $H_\alpha(p)$ whenever $m>0$. On the unimodular crossing branch $X=-1$, the deficit $\Delta_\alpha$ therefore simplifies to $H_\alpha(p)+H_\alpha(q)-\log r$. Because the Abelian modular matrix has flat entries $|S_{ab}|=r^{-1/2}$, combining R\'enyi monotonicity in $\alpha$ with the Maassen--Uffink uncertainty relation~\cite{MaassenUffink1988} gives
\begin{equation}
\begin{gathered}
H_\alpha(p)+H_\alpha(q)\geq H(p)+H(q)\geq\log r
 \\(0<\alpha\leq1).
\end{gathered}
\label{eq:pointed-positive}
\end{equation}
When $X\ge0$ and the input regions carry distinct essential slopes, neither output contains an $E$ component ($m_U=m_I=0$) and $\Delta_\alpha$ is a sum of nonnegative terms. When $X\ge0$ and all state-bearing regions share a common slope, the deficit reduces to $2X\log D+\kappa H_\alpha(p)$, with $\kappa$ defined in Eq.~\eqref{eq:common-counts}. As shown above, the only negative value of $\kappa$ is $\kappa=-1$, which occurs when $m_A=m_B=0$ and exactly one output has $m=k\ge1$. The homological bound $X+\Delta m\ge0$ then ensures $X\ge k\ge1$, and since $H_\alpha(p)\le\log r=2\log D$, the topological contribution $2X\log D$ suffices to offset $-H_\alpha(p)$. Together with Eq.~\eqref{eq:pointed-positive}, this exhaustion of geometric branches proves universal strong subadditivity for every pointed category throughout $0<\alpha\le1$.

For orders $\alpha>1$, two single annuli crossing once realize $X=-1$ and evaluate $\Delta_\alpha$ directly to the entropic Fourier residual in Eq.~\eqref{eq:pointed-high}, making that condition necessary for universal positivity. Conversely, in every $X\ge0$ branch the pointed deficit depends only on the nonnegativity of $H_\alpha(p)$ and the upper bound $H_\alpha(p)\le\log r$, both of which hold for all $\alpha>1$. Condition~\eqref{eq:pointed-high} is therefore both necessary and sufficient, completing the proof of Theorem~\ref{thm:classification}. As the explicit pointed counterexamples in Section~\ref{sec:mechanism} demonstrate, a violation of strong subadditivity at $\alpha>1$ can arise purely from this crossing-basis Fourier residual and does not by itself diagnose non-Abelian anyons.

%% file: sections/microscopic.tex
\section{Microscopic realization in quantum-double ground states}\label{sec:microscopic}
\label{sm:micro}
Quantum-double and string-net models provide explicit lattice realizations of doubled topological phases~\cite{Kitaev2003FaultTolerantAnyons,LevinWen2005StringNet}, in which exact model mappings and projected entangled pair state (PEPS) representations link microscopic link variables to topological sectors~\cite{BuerschaperAguado2009Mapping,SchuchCiracPerezGarcia2010PEPS}. Torus entanglement entropies in the toric code and reduced spectra for string-net states serve as standard lattice benchmarks~\cite{HammaIonicioiuZanardi2005Kitaev,LuoHuWu2016TorusEntanglement}. Here we extend the one- and two-annulus analysis of Ref.~\cite{Liu2024QuantumDimensions} for a specified boundary-holonomy ground state to any even number of alternating strips that fit on the lattice. This generalization covers the three-annulus geometry needed to compute the four-region R\'enyi deficit directly from physical edge degrees of freedom.

\subsection{State and fixed spatial tensor factors}
Subsystem entropies in lattice gauge theories and string-net models depend on the chosen local operator algebra or Hilbert-space factorization~\cite{CasiniHuertaRosabal2014GaugeEntropy}. In non-Abelian link spaces, the entropy combines a Shannon term over superselection sectors, representation-dimension weights, and entanglement within each sector~\cite{Donnelly2012GaugeEntropy}, whereas operational protocols restricted to gauge-invariant local operations extract only the distillable component~\cite{VanAcoleyenEtAl2016Distillation}. A parallel gauge-theoretic formulation based on the tube algebra applies to Levin--Wen models~\cite{KawagoeEtAl2024LevinWenGauge}. Throughout this section, we work in the full physical edge tensor product and keep that single factorization fixed across all four cuts entering the deficit.

Let $G$ be a finite group of order $q$. On a connected cellulation of the torus, the untwisted quantum-double model assigns a local Hilbert space $\mathbb C[G]$ to each directed edge $e$, forming the total space $\bigotimes_e\mathbb C[G]$, and defines the ground space through commuting vertex gauge projectors and face-flatness projectors. Choosing a transverse cycle $x$, we examine the state
\begin{equation}
 |\Omega_\perp\rangle=|\mathscr F_x|^{-1/2}\sum_{g\in\mathscr F_x}|g\rangle,
 \quad \mathscr F_x=\{g:\text{all faces flat},\ \operatorname{Hol}_x(g)=1\}.
 \label{sm:state}
\end{equation}
Vertex gauge transformations permute $\mathscr F_x$ while preserving the trivial holonomy along $x$, so $|\Omega_\perp\rangle$ is a normalized pure ground state whose longitudinal holonomy remains unconstrained.

To evaluate distinct regions within a single Hilbert-space factorization, we install all candidate interfaces on one common lattice refinement before specifying any subsystem. Following Ref.~\cite{Liu2024QuantumDimensions}, we duplicate each interface edge into two identically oriented parallel edges that bound a flat bigon. The isometric embedding $|g\rangle\mapsto|g,g\rangle$ intertwines the vertex gauge action and maps $|\Omega_\perp\rangle$ into the ground space of the refined cellulation. The two copies of every duplicated interface edge are assigned to the two adjacent elementary slabs, and all remaining edges of the cellulation are partitioned among those slabs. Each subsystem is then specified as a union of disjoint physical edge sets, with its reduced density matrix defined by the standard partial trace over the complementary edge factors. When an installed seam lies wholly inside one subsystem rather than along the active cut, the local unitary $|g,h\rangle\mapsto|g,g^{-1}h\rangle$ acts within that subsystem to map the bigon pair $|g,g\rangle$ to $|g,1\rangle$, which decouples the second register. Changing the cut therefore alters only the local basis transformations inside the subsystems while leaving the state, the refined lattice, and the physical edge tensor factors fixed.

\begin{theorem}[Ordinary-edge spectrum and spatial violations]\label{thm:microscopic}
Let $G$ be a finite group of order $q$, and let $|\Omega_\perp\rangle$ be the ground state in Eq.~\eqref{sm:state} on a common torus refinement carrying all candidate interfaces. Consider any union $R_m$ of $m\ge1$ alternating longitudinal annuli, with the fixed transverse cycle $x$ traversing the $2m$ active cylinders once in cyclic order. Suppose these cylinders fit on the refined lattice and each contain at least one interior vertex. Let $B_R$ denote the total number of active boundary vertices and set $p_a=d_a^2/q^2$ for $a\in\operatorname{Irr}D(G)$. The ordinary reduced density matrix $\rho_{R_m}$ on the physical edge tensor factor has nonzero eigenvalues $\lambda_a$ with multiplicities $r_a$ given by
\begin{equation}
 Q_R=q^{B_R-2m},\qquad \lambda_a=\frac{p_a}{Q_Rd_a^{2m}},\qquad r_a=Q_Rd_a^{2m}.
 \label{sm:full-spectrum}
\end{equation}
When the common refinement accommodates the two fixed region pairs $(A_\mp,B_\mp)$ of Eqs.~\eqref{eq:regions-low} and~\eqref{eq:regions-high}, their four-region microscopic R\'enyi deficits take the exact form in Eq.~\eqref{sm:exact-deficits}. If $G$ is non-Abelian, the low-order pair satisfies $\Delta_\alpha^-<0$ for all $0<\alpha<1$ and the high-order pair satisfies $\Delta_\alpha^+<0$ for all finite $\alpha>1$, whereas $\Delta_1^\mp=0$; both signs are achieved by the same state $|\Omega_\perp\rangle$ and the two $\alpha$-independent region pairs. If $G$ is Abelian, $\Delta_\alpha^\mp=0$ for every finite $\alpha>0$.
\end{theorem}
We prove Theorem~\ref{thm:microscopic} in three steps: we pass from physical edges to strip holonomy coordinates, isolate the unentangled registers and boundary Bell pairs from the multi-strip core, and then perform the Fourier decomposition to evaluate the two fixed region pairs.

\subsection{From physical edges to strip coordinates}
After the inactive seam registers are decoupled into product states, the active cuts divide the torus into $n=2m$ alternating cylinders $T_i$. Each cylinder $T_i$ meets its incoming and outgoing active boundaries along $L_{i-1}$ and $L_i$ vertices, respectively, and contains $N_i\ge1$ interior vertices, where vertices on inactive seams are counted in the interior. Because edge duplication adds no vertices, we write $V_i=N_i+L_{i-1}+L_i$, $N=\sum_iN_i$, and $B_R=\sum_iL_i$, giving $V=N+B_R$ vertices in total on the torus.

\input{sections/figure-2.tex}

In each cylinder $T_i$, we choose an interior root vertex $b_i$, a spanning tree of rooted paths $P_{i,v}$ with group holonomies $t_{i,v}$ satisfying $t_{i,b_i}=1$, and a positively oriented longitudinal loop based at $b_i$ with holonomy $k_i$. Group multiplication along a path places the last traversed edge on the left, while edges traversed against their orientation carry inverses. For any directed edge $e:v\to w$ in $T_i$, let $\nu_{i,e}\in\mathbb Z$ be the longitudinal winding number of the closed loop $P_{i,v}eP_{i,w}^{-1}$. Face flatness fixes the edge variable uniquely:
\begin{equation}
 U_e=t_{i,w}k_i^{\nu_{i,e}}t_{i,v}^{-1}.
 \label{sm:edge-reconstruction}
\end{equation}
Conversely, every assignment of $k_i\in G$ and of the $V_i-1$ tree holonomies $t_{i,v}$ ($v\ne b_i$) specifies a flat edge configuration on $T_i$. Path products along the tree telescope, contractible cycles have zero net winding, and transport around the chosen longitudinal generator returns $k_i$ because every based loop on the cylinder is homotopic to an integer power of that generator. Equation~\eqref{sm:edge-reconstruction} thus defines a bijection between flat configurations on $T_i$ and $q^{V_i}$ group tuples, with $t_{i,b_i}=1$ fixing the trivial path at the root. Because this map merely reparametrizes the physical edge basis rather than taking a gauge quotient, mutually conjugate holonomies remain distinct configurations.

We now align the basepoints and orientations of all active boundary circles. On each cylinder $T_i$, we select the spanning tree so that it runs along the open boundary arcs and joins them to the root $b_i$ via transverse paths, leaving a single closing edge of unit longitudinal winding on each boundary circle. Concatenating these transverse paths across all $n$ cylinders produces the global cycle $x$. Write $g_{i,r}$ and $h_{i,r}$ for the tree holonomies to the $r$-th vertex on the incoming and outgoing boundaries of $T_i$, respectively. Because the duplicated open boundary edges carry equal group elements across the seam between $T_i$ and $T_{i+1}$, the ratio $a_i=g_{i+1,r}^{-1}h_{i,r}$ is independent of the vertex index $r$. Matching the closing boundary edges and imposing trivial holonomy along $x$ then gives the inter-strip sewing constraints
\begin{equation}
 h_{i,r}=g_{i+1,r}a_i,\qquad k_{i+1}=a_i k_i a_i^{-1},\qquad a_n\cdots a_1=1.
 \label{sm:strip-constraints}
\end{equation}
Here cylinder indices are read modulo $n$. Any configuration in $\mathscr F_x$ is uniquely determined by freely specifying the $B_R$ incoming boundary holonomies $g_{i,r}$, the $N-n$ nonroot interior holonomies, the transfer variables $a_1,\ldots,a_{n-1}$, and the initial longitudinal holonomy $k_1$. The final transfer element is fixed by $a_n=(a_{n-1}\cdots a_1)^{-1}$, and the remaining holonomies $h_{i,r}$ and $k_i$ follow from Eq.~\eqref{sm:strip-constraints} in a way that automatically respects periodic closure. The configuration set therefore has cardinality
\begin{equation}
 |\mathscr F_x|=q^{B_R+(N-n)+(n-1)+1}=q^{B_R+N}=q^V.
 \label{sm:configuration-count}
\end{equation}
Accordingly, every configuration in $|\Omega_\perp\rangle$ enters with the same amplitude $q^{-V/2}$, independent of the active-strip partition.

\subsection{Local pure factors and the entangled core}
On a cylinder $T_i$ with $E_i$ edges, the reconstruction map~\eqref{sm:edge-reconstruction} stores the $q^{V_i}$ flat configurations in $V_i$ group registers and sets the remaining $E_i-V_i\ge0$ redundant edge registers to $|1\rangle$. Extending this map to an orthonormal basis of the full $E_i$-edge Hilbert space yields a strip-local unitary transformation that preserves all physical degrees of freedom. On the $V_i$ active registers of $T_i$, we introduce the invertible group-coordinate transformation
\begin{align}
 q_i&=g_{i,1},&x_i&=h_{i,1}g_{i,1}^{-1},&y_i&=g_{i,1}k_i g_{i,1}^{-1},\nonumber\\
 u_{i,r}&=g_{i,r}g_{i,1}^{-1},&v_{i,r}&=h_{i,r}h_{i,1}^{-1}&&(r>1).
 \label{sm:coordinates}
\end{align}
Its inverse is $g_{i,1}=q_i$, $h_{i,1}=x_iq_i$, $k_i=q_i^{-1}y_iq_i$, $g_{i,r}=u_{i,r}q_i$, and $h_{i,r}=v_{i,r}x_iq_i$. Because this transformation acts entirely within a single strip $T_i$, and each $T_i$ lies wholly inside either $R$ or $\bar R$, no finite-depth circuit restriction is required. In these new coordinates, the sewing relations~\eqref{sm:strip-constraints} reduce to
\begin{equation}
 v_{i,r}=u_{i+1,r},\qquad y_{i+1}=x_i y_i x_i^{-1},\qquad x_n\cdots x_1=1.
 \label{sm:core-constraints}
\end{equation}
Both the reference variables $q_i$ and the $N-n$ nonroot interior holonomies drop out of Eq.~\eqref{sm:core-constraints}, so they factor out as $N$ unentangled registers in the uniform superposition $|+_q\rangle=q^{-1/2}\sum_g|g\rangle$. For every $r>1$, the relation $v_{i,r}=u_{i+1,r}$ locks an outgoing register of $T_i$ to an incoming register of $T_{i+1}$ in a maximally entangled state of dimension $q$, yielding $B_R-n$ boundary Bell pairs across the active seams. The remaining $2n$ registers $\{x_i,y_i\}_{i=1}^n$ carry the $n$-strip entangled core
\begin{equation}
 |\Phi_n\rangle=q^{-n/2}\sum_{\substack{\{x_i,y_i\}:\ y_{i+1}=x_i y_i x_i^{-1}\\x_n\cdots x_1=1}}\bigotimes_{i=1}^n|x_i,y_i\rangle.
 \label{sm:core-state}
\end{equation}
Freely choosing $y_1$ and $x_1,\ldots,x_{n-1}$ fixes the remaining core variables uniquely, so the sum has $q^n$ terms. Combining the normalization factors of the $N$ pure registers, the $B_R-n$ Bell pairs, and the core state reproduces $q^{-N/2}q^{-(B_R-n)/2}q^{-n/2}=q^{-V/2}$, in agreement with Eq.~\eqref{sm:configuration-count}. Because each strip $T_i$ lies on one side of the bipartition, the total basis change factors as $W_R\otimes W_{\bar R}$ and preserves the entanglement spectrum of $\rho_R$; switching between cuts merely changes these strip-local unitaries.

\subsection{Fourier decomposition and reduced spectrum}
In every term of $|\Phi_n\rangle$, the conjugation chain $y_{i+1}=x_i y_i x_i^{-1}$ confines all $n$ longitudinal holonomies $y_i$ to a single conjugacy class $C\subset G$. For each class $C$, we pick a representative $c\in C$ and fix coset representatives $t_y\in G$ satisfying $t_yct_y^{-1}=y$ for all $y\in C$. The elements
\begin{equation}
 z_i=t_{y_{i+1}}^{-1}x_it_{y_i}\in Z_c,\qquad z_n\cdots z_1=1,
 \label{sm:centralizer}
\end{equation}
then lie in the centralizer $Z_c=\{z\in G:zc=cz\}$ and obey a cyclic product constraint, which puts the allowed strip pairs $(x_i,y_i)$ satisfying $y_{i+1}=x_i y_i x_i^{-1}$ in bijection with tuples $(C,y_{i+1},y_i,z_i)$. Using the unitary Peter--Weyl matrix elements $\sqrt{d_\mu/|Z_c|}D^\mu_{ab}(z)$ for the irreducible representations $\mu$ of $Z_c$ of dimension $d_\mu$, we expand the delta function on $Z_c$ as
\begin{equation}
 \delta(z_n\cdots z_1,1)=\frac1{|Z_c|}\sum_\mu d_\mu\,\overline{\operatorname{Tr}[D^\mu(z_n)\cdots D^\mu(z_1)]}.
 \label{sm:character}
\end{equation}
Performing the inverse Fourier transform on each strip and applying Schur orthogonality locks the representation label $\mu$ across all $n$ strips while contracting the matrix indices cyclically around the torus. Combining the character prefactor in Eq.~\eqref{sm:character}, the $n$ unitary Fourier normalizations, and the $n$ index contractions yields the net factor
\begin{equation}
 \frac{d_\mu}{|Z_c|}\left(\frac{d_\mu}{|Z_c|}\right)^{n/2}\left(\frac{|Z_c|}{d_\mu}\right)^n
 =\left(\frac{|Z_c|}{d_\mu}\right)^{n/2-1}.
 \label{sm:schur-factor}
\end{equation}
Irreducible representations of the quantum double $D(G)$ are labeled by pairs $a=(C,\mu)$ of quantum dimension $d_a=|C|d_\mu$. Grouping the class element $y\in C$ and the representation index of $\mu$ into a composite seam index $J=1,\ldots,d_a$, and combining $q=|C||Z_c|$ with the core normalization $q^{-n/2}$, we obtain the sector decomposition
\begin{align}
 |\Phi_n\rangle&=\sum_a\frac{d_a}{q}|\mathrm{Ring}_{a,n}\rangle,\nonumber\\
 |\mathrm{Ring}_{a,n}\rangle&=d_a^{-n/2}\sum_{J_1,\ldots,J_n=1}^{d_a}\bigotimes_{i=1}^n|a;J_{i+1},J_i\rangle.
 \label{sm:ring}
\end{align}
Each $|\mathrm{Ring}_{a,n}\rangle$ is a normalized state formed by $n$ maximally entangled bonds of dimension $d_a$ connecting adjacent strips, and the local strip states $|a;J_{i+1},J_i\rangle$ for distinct anyon sectors $a$ are mutually orthogonal on every strip.

For the alternating union $R_m$ of $m$ disjoint annuli ($n=2m$), all $2m$ bonds of $|\mathrm{Ring}_{a,n}\rangle$ traverse the active boundary between $R_m$ and $\bar R_m$. Tracing out $\bar R_m$ therefore removes all off-diagonal coherences between distinct sectors $a$. Inside each sector $a$, the reduced ring state is maximally mixed of rank $d_a^{2m}$ and enters the reduced core with probability $p_a=d_a^2/q^2$. Restoring the $B_R-2m$ boundary Bell pairs of dimension $q$ establishes the nonzero entanglement spectrum in Eq.~\eqref{sm:full-spectrum}.
Because $\sum_a r_a\lambda_a=\sum_ap_a=1$, the $\alpha$-th spectral moment follows immediately:
\begin{equation}
 \operatorname{Tr}\rho_R^\alpha=Q_R^{1-\alpha}\sum_a p_a^\alpha d_a^{2m(1-\alpha)}.
 \label{sm:spectral-moment}
\end{equation}
For every finite R\'enyi order $\alpha>0$ with $\alpha\ne1$, this trace yields the microscopic entanglement entropy of $R_m$ directly, without replica analytic continuation:
\begin{equation}
 S_\alpha^{\mathrm{micro}}(R_m)=B_R\log q-2m\log q+\frac1{1-\alpha}\log\sum_a p_a^\alpha d_a^{2m(1-\alpha)}.
 \label{sm:micro-entropy}
\end{equation}
\begin{equation}
 S_\alpha^{\rm micro}(R_m)=B_R\log q+S_\alpha^{\rm top}(R_m;p).
 \label{eq:microscopic-dictionary}
\end{equation}
At $m=1$, the topological terms cancel to leave $S_\alpha^{\mathrm{micro}}(R_1)=B_R\log q$; at $m=2$, the subleading term $(1-\alpha)^{-1}\log\sum_a(d_a/q)^{4-2\alpha}$ recovers the two-annulus result of Ref.~\cite{Liu2024QuantumDimensions}. At $m=3$, the same ground state produces a subleading sum with exponent $6-4\alpha$. Sending $\alpha\to1$ gives the von Neumann entropy
\begin{equation}
 S_1^{\mathrm{micro}}(R_m)=B_R\log q-2(m-1)\left(\log q-\sum_a p_a\log d_a\right).
 \label{sm:micro-von-neumann}
\end{equation}
Here the topological correction is affine in the component count $m$. Equations~\eqref{sm:full-spectrum}--\eqref{sm:micro-von-neumann} apply to every $m\ge1$ whose $2m$ alternating strips fit on the refined torus cellulation.

\subsection{Two fixed region pairs for every non-Abelian finite group}
To realize the two curvature regimes on the lattice, we parametrize the transverse circle by $[0,12)$ and write $U=A\cup B$ and $I=A\cap B$. The low-order pair $(A_-,B_-)$ is defined by $A_-=([0,2]\cup[3,5])\times S^1$ and $B_-=([1,4]\cup[4.5,6])\times S^1$, which has annulus counts $(m_A,m_B;m_U,m_I)=(2,2;1,3)$. The high-order pair $(A_+,B_+)$ is defined by $A_+=[1,5]\times S^1$ and $B_+=([0,2]\cup[4,6]\cup[8,9])\times S^1$, which has annulus counts $(m_A,m_B;m_U,m_I)=(1,3;2,2)$. A single half-integer transverse grid contains all interface seams of both pairs; installing these seams on one common lattice refinement fixes the physical edge subsets for $A_\pm$, $B_\pm$, and their set-theoretic unions and intersections.

In each pair, $A$ and $B$ have disjoint boundaries, so every active seam of $A$ or $B$ appears once in the boundary of $U$ or $I$. Consequently, the area-law contributions $L_i\log q$ cancel seam by seam in the four-region combination
\begin{equation}
 \Delta_\alpha^{\mathrm{micro}}(A,B)=S_\alpha^{\mathrm{micro}}(A)+S_\alpha^{\mathrm{micro}}(B)-S_\alpha^{\mathrm{micro}}(U)-S_\alpha^{\mathrm{micro}}(I).
 \label{sm:micro-deficit}
\end{equation}
This seam-wise cancellation holds even when the individual boundary circles carry different lattice lengths $L_i$. Because both configurations satisfy $m_A+m_B=m_U+m_I=4$, the $-2m\log q$ terms in Eq.~\eqref{sm:micro-entropy} also cancel. Setting $T_G(s)=\sum_{a\in\operatorname{Irr}D(G)}d_a^s$ and using $T_G(2)=q^2$, we obtain the exact microscopic deficits
\begin{equation}
 \Delta_\alpha^-=\frac1{1-\alpha}\log\frac{T_G(4-2\alpha)^2}{q^2T_G(6-4\alpha)},\qquad \Delta_\alpha^+=-\Delta_\alpha^-.
 \label{sm:exact-deficits}
\end{equation}
The superscripts $\mp$ refer to $(A_\mp,B_\mp)$. Any non-Abelian group $G$ has at least one noncentral conjugacy class; combining that class with the trivial centralizer representation gives an anyon sector of dimension $d_a>1$ alongside the vacuum sector $d_0=1$. Because both sectors appear with positive weight in $T_G(s)$, the logarithmic moment-generating function has strictly positive second derivative:
\begin{equation}
 \frac{d^2}{ds^2}\log T_G(s)=\operatorname{Var}_{v(s)}(\log d_a)>0,
 \qquad v_a(s)=\frac{d_a^s}{T_G(s)}.
 \label{sm:group-convexity}
\end{equation}
Because $4-2\alpha$ is the midpoint between $2$ and $6-4\alpha$, strict log-convexity of $T_G(s)$ forces $\Delta_\alpha^-<0$ for all $0<\alpha<1$ and $\Delta_\alpha^+<0$ for all finite $\alpha>1$, whereas $\Delta_1^\mp=0$. Both the state $|\Omega_\perp\rangle$ and the two region pairs are chosen independently of $\alpha$; when $G$ is Abelian, every sector has $d_a=1$, so $T_G(s)=q^2$ and $\Delta_\alpha^\mp=0$ for all $\alpha>0$.

For the simplest non-Abelian group $G=S_3$ ($q=6$), the eight irreducible representations of $D(S_3)$ have quantum dimensions $(1,1,2,3,3,2,2,2)$, so $T_{S_3}(s)=2+4\cdot2^s+2\cdot3^s$. For $m=1,2,3$, the reduced core states have ranks $\sum_a d_a^{2m}=36,228,1716$ and sector eigenvalues $1/36$, $1/(36d_a^2)$, and $1/(36d_a^4)$, respectively. Evaluating Eq.~\eqref{sm:exact-deficits} at $\alpha=1/2$ on $(A_-,B_-)$ and at $\alpha=2$ on $(A_+,B_+)$ gives the explicit negative values
\begin{equation}
\begin{gathered}
\Delta_{1/2}^{\rm micro}(A_-,B_-)=2\log\frac{484}{513}<0,\\
 \Delta_2^{\rm micro}(A_+,B_+)=\log\frac{16}{29}<0.
\end{gathered}
\label{eq:microscopic-values}
\end{equation}
Under standard partial traces over the fixed physical-edge tensor factors, the ground state $|\Omega_\perp\rangle$ and the region pairs $(A_\mp,B_\mp)$ therefore exhibit the shared-flux violations directly on the lattice. Section~\ref{sec:filtered-witness} establishes a controlled continuation of the R\'enyi-two witness for every fixed finite non-Abelian group. Its deficit remains strictly negative in the stated joint system-size and interface-separation limit for sufficiently weak filters from the bounded diagonal class defined there, applied to these same physical edges.

%% file: sections/figure-2.tex
\begin{figure}[!p]
 \centering
 \includegraphics[width=\linewidth]{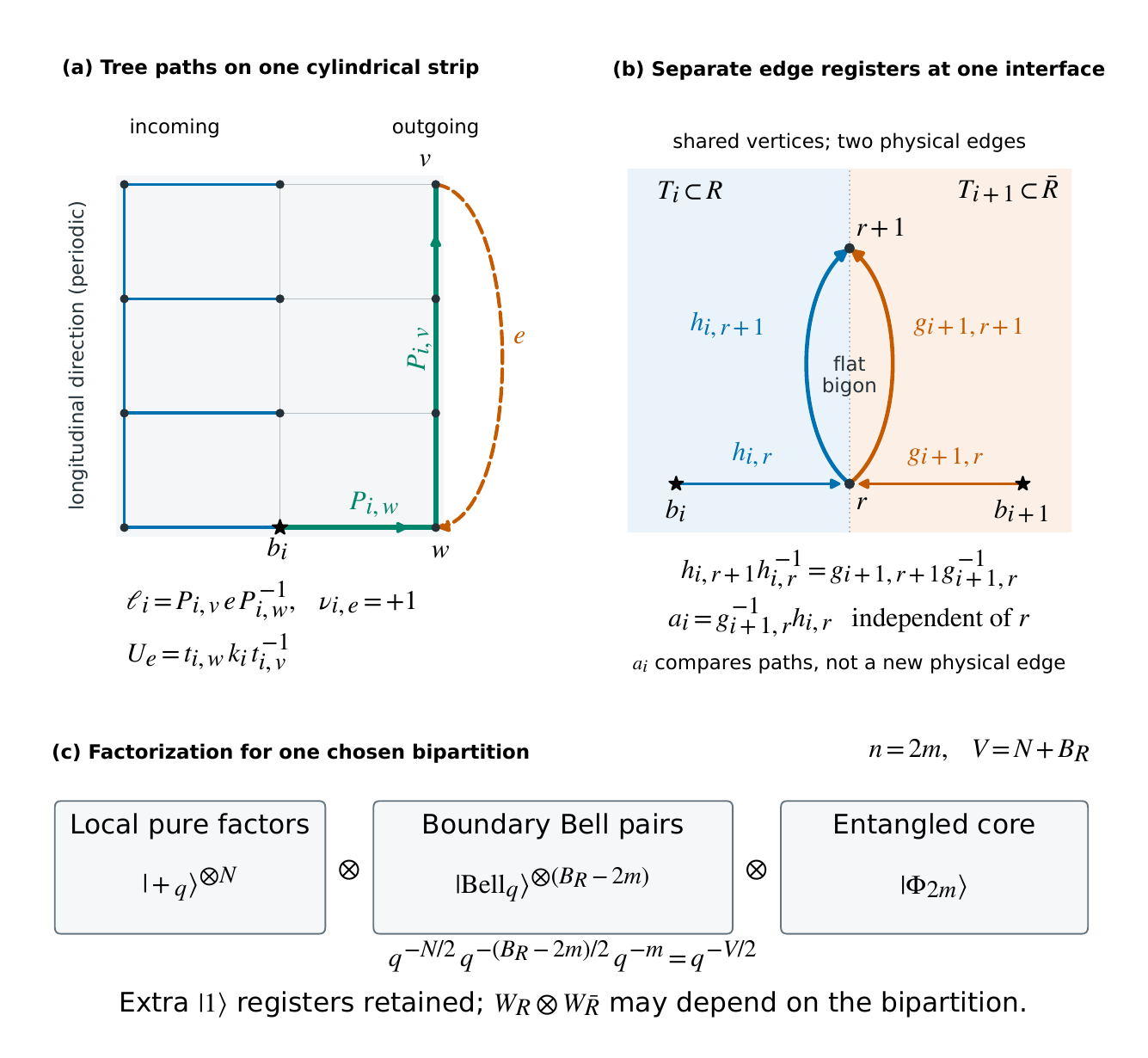}
 \caption{Microscopic strip coordinates and interface factorization. (a) A spanning tree (medium and thick solid lines) on a cylindrical strip $T_i$ with interior root $b_i$; thin gray lines show non-tree edges. The highlighted tree paths $P_{i,v}$ and $P_{i,w}$ (thick solid lines) connect the root to vertices $v$ and $w$. The longitudinal direction is periodic. A single closing edge $e:v\to w$ (dashed line) is drawn, carrying longitudinal winding $+1$; the loop $P_{i,v}eP_{i,w}^{-1}$ serves as the based generator $\ell_i$ with holonomy $k_i$. Remaining closing links are omitted. Setting $t_{i,b_i}=1$ reflects the empty path holonomy at the root rather than a gauge quotient. (b) An active open interface segment shared between adjacent strips $T_i\subset R$ and $T_{i+1}\subset\bar R$. Physical vertices are shared, whereas the two co-oriented edge copies belong to separate strip registers across a flat bigon. The variables $g$ and $h$ represent root-to-vertex path holonomies, not single-edge labels. Enforcing equal edge variables across the bigon makes $a_i=g_{i+1,r}^{-1}h_{i,r}$ independent of $r$, while matching the closing links supplies the conjugation relation $k_{i+1}=a_ik_ia_i^{-1}$ in Eq.~\eqref{sm:strip-constraints}. (c) Tensor-factor accounting for a chosen alternating $2m$-strip partition. Strip-local basis transformations isolate $N$ uniform single-register states, $B_R-2m$ boundary Bell pairs, and the entangled core $|\Phi_{2m}\rangle$, retaining spectator registers at $|1\rangle$. The product normalization $q^{-N/2}q^{-(B_R-2m)/2}q^{-m}=q^{-V/2}$ recovers the global amplitude on $V=N+B_R$ vertices. Microscopic lattice edges, the physical state, and the global Hilbert-space factorization are invariant, while the calculational unitaries are chosen per bipartition.}
 \label{fig:strip-coordinate-interface}
\end{figure}

Figure~\ref{fig:strip-coordinate-interface} illustrates the spanning-tree parameterization of an individual strip, the edge matching across adjacent interfaces, and the resulting tensor-factor decoupling.

%% file: sections/deformation.tex
\section{Controlled deformation beyond the fixed point}
\label{sec:filtered-witness}

At the exact fixed point, every finite non-Abelian group yields a strictly negative R\'enyi-two deficit. We now show that this negative value persists in the thermodynamic limit when physical-edge diagonal filters alter the regional entanglement spectra at finite volume. Our analysis combines three distinct elements: a connected-cluster expansion that bounds the four-region purity ratio uniformly across a bounded filter class, a finite-volume Taylor calculation that establishes genuine spectral variation on graphs free of extra short cycles, and an independent local-Hamiltonian stability argument that embeds the filtered reference state into a uniformly gapped ground space.

\subsection{Filtered state and main result}
Wavefunctions with finite correlation lengths, including deformed toric-code states, are widely used to analyze phase transitions and subleading geometric contributions to entanglement entropy~\cite{CastelnovoChamon2008DeformedToric,PapanikolaouRamanFradkin2007FiniteCorrelation}. Applying local physical filters to fixed-point tensor networks provides a standard mechanism for generating such deformations~\cite{HaegemanEtAl2015Shadows}. For the von Neumann topological entanglement entropy, several techniques control deviations from the fixed point, ranging from first-order perturbation theory~\cite{Kim2012PerturbativeTEE} and reference-state circuit bounds~\cite{KimLevinLin2023EtAl} to lower bounds based on sector and fusion rules~\cite{Levin2024TEEBound}. R\'enyi entropies are much more delicate: geometric boundary cancellation alone does not prevent spurious contributions in subsystem-symmetry-protected phases or boundary-state models~\cite{WilliamsonDuaCheng2019,KatoBrandao2020SpuriousBoundary}, and replica correlation lengths can remain large even when physical correlations are short-ranged~\cite{ZouHaah2016}. Rather than averaging purities over shallow random circuits as in studies of cyclic-group doubles~\cite{OlivieroLeoneZhouHamma2022TopologicalPurity}, we directly bound the deterministic four-region purity ratio of a single filtered state.

Let $G$ be a fixed finite non-Abelian group of order $q$, and choose a real cost function $b:G\to[0,1]$. For a real deformation parameter $\lambda$, we act with the identical single-edge diagonal filter on every physical edge of the common refined state from Theorem~\ref{thm:microscopic}:
\begin{equation}
\begin{gathered}
f_\lambda|g\rangle=e^{-\lambda b(g)}|g\rangle,\qquad
F_\lambda=\bigotimes_{e\ {\mathrm{physical}}}f_\lambda,\\
|\Psi_\lambda\rangle=
\frac{F_\lambda|\Omega_\perp\rangle}{\|F_\lambda|\Omega_\perp\rangle\|}.
\end{gathered}
\label{eq:sm-filter-state}
\end{equation}
Every duplicated seam—whether active or dormant—carries two identically oriented physical edges, and both copies are weighted by $f_\lambda$. All four regional density matrices are obtained from this single normalized state $|\Psi_\lambda\rangle$ by taking standard partial traces over the fixed physical-edge tensor product.

We place the common refinement on an $N_x\times L$ rectangular torus ($N_x,L\ge3$) with $V=N_xL$ vertices and maximum physical multigraph degree $\mathfrak d\le6$, counting parallel edges separately. The lattice is an integral dilation chosen large enough that every active cylinder contains at least one interior vertex. We retain the high-order region pair $(A_+,B_+)$ together with its exact set-theoretic union $U=A_+\cup B_+$ and intersection $I=A_+\cap B_+$. Let $W$ denote the minimum graph distance between distinct boundary circles in the common pool $\partial A_+\cup\partial B_+$, and define the characteristic length scale $s_*=\min(W,L)$. Writing $r_G=|\operatorname{Irr}D(G)|$ and recalling the moment function $T_G(s)$ from Eq.~\eqref{sm:exact-deficits}, we denote the exact fixed-point R\'enyi-two deficit by
\begin{equation}
 \delta_G=\log\frac{r_G^2}{q^2T_G(-2)}<0.
 \label{eq:group-benchmark}
\end{equation}

\begin{theorem}[Controlled diagonal deformation]\label{thm:filtered}
Under the state, lattice, and region specifications above, the following properties hold:
\begin{enumerate}
\item For any function $b:G\to[0,1]$ and any real parameter $|\lambda|\le\lambda_{\rm CE}$, the microscopic R\'enyi-two deficit satisfies
\begin{equation}
\begin{gathered}
\lambda_{\rm CE}=\frac14\log\left(1+\frac{1}{2e^3\mathfrak d^2}\right),\\
\left|\Delta_2^{\rm micro}(\lambda;N_x,L)-\delta_G\right|
\le 4N_xL\,e^{-s_*}.
\end{gathered}
\label{eq:sm-filter-bound}
\end{equation}
In particular, $\Delta_2^{\rm micro}(\lambda;N_x,L)$ converges to $\delta_G$ uniformly over all $b:G\to[0,1]$ along any lattice sequence satisfying $N_xL e^{-\min(W,L)}\to0$.
\item Fix an aspect-ratio range bounded away from zero and infinity. For each fixed group $G$, one can choose a threshold $\lambda_{\rm gap}(G)>0$, a minimum linear size, and a positive excitation-gap bound independently of $b:G\to[0,1]$ and the torus dimensions. On every such torus above that minimum size and for every $|\lambda|\le\lambda_{\rm gap}(G)$, the state $|\Psi_\lambda\rangle$ lies in the zero-energy ground space of the local parent Hamiltonian in Eq.~\eqref{eq:sm-filter-parent}, with the stated gap above that space. The notation $\lambda_{\rm gap}(G)$ leaves the fixed aspect-ratio bound implicit.
\item On the original straight-strip graph, if $L,W\ge3$ and the installed seam bigons are the only two-edge cycles, then for each region $R\in\{A_+,\allowbreak B_+,\allowbreak U,\allowbreak I\}$ the purity $P_R(\lambda)=\operatorname{Tr}\rho_R(\lambda)^2$ obeys
\begin{equation}
 (\log P_R)'(0)=0,\qquad
 (\log P_R)''(0)=32\operatorname{Var}_{\rm Haar}(b)B_R.
 \label{eq:sm-filter-spectrum-change}
\end{equation}
Here $\operatorname{Var}_{\rm Haar}(b)$ is the variance with respect to the uniform Haar measure on $G$. For each fixed nonconstant $b$ and every fixed finite lattice satisfying these graph conditions, the regional purity is nonconstant near $\lambda=0$, so the regional entanglement spectrum changes.
\end{enumerate}
\end{theorem}

We establish the cluster bound, the finite-volume curvature identity, and the parent-Hamiltonian gap in separate subsections below. When $|\lambda|\le\min(\lambda_{\rm CE},\lambda_{\rm gap}(G))$ and the lattice dimensions satisfy $4N_xLe^{-s_*}\le-\delta_G/2$, the deformed state simultaneously enjoys a uniformly gapped local parent and a strictly negative deficit $\Delta_2^{\rm micro}\le\delta_G/2<0$. Scaling a bounded-aspect-ratio torus together with the cut geometry achieves $N_xLe^{-s_*}\to0$. Neither threshold represents a physical phase transition: $\lambda_{\rm CE}\simeq1.72812\times10^{-4}$ (at $\mathfrak d=6$) is an explicit polymer convergence radius, while $\lambda_{\rm gap}(G)$ is an existential stability radius uniform in $b$ at fixed $G$, with no claim of uniformity across groups of growing order.

\subsection{Exact replica constraints on the physical edges}
Assign each flat configuration $g\in\mathscr F_x$ the product weight $w_\lambda(g)=\prod_e e^{-\lambda b(g_e)}$. For a subsystem $R$, let $\chi_R(g,h)$ denote the indicator that equals $1$ precisely when exchanging the edge labels inside $R$ between $g$ and $h$ produces two configurations that both remain in $\mathscr F_x$. The state normalization and regional purity are then given by
\begin{equation}
\begin{gathered}
Z_\lambda=\sum_{g\in\mathscr F_x}w_\lambda(g)^2,\qquad
P_R(\lambda)=\frac{Z_R^{(2)}(\lambda)}{Z_\lambda^2},\\
Z_R^{(2)}(\lambda)=\sum_{g,h\in\mathscr F_x}
 w_\lambda(g)^2w_\lambda(h)^2\chi_R(g,h).
\end{gathered}
\label{eq:sm-filter-replica}
\end{equation}
Because the swap operation merely permutes the two group labels on each edge inside $R$ and leaves the labels outside $R$ untouched, the unordered pair of edge variables on every physical edge is preserved. For real diagonal filters, the product of the four wavefunction weights therefore collapses to $w_\lambda(g)^2w_\lambda(h)^2$ without assuming that $b$ is conjugation-invariant. When we form the four-region combination $P_AP_B/(P_UP_I)$, the overall normalization factors $Z_\lambda$ cancel out.

To resolve the flatness constraint $g\in\mathscr F_x$, we choose a fixed integer longitudinal cocycle $\nu_e\in\{0,1,-1\}$ and write every physical edge variable in terms of vertex frames $u_x\in G$ and a longitudinal holonomy $k\in G$:
\begin{equation}
 U_e(u,k)=u_y k^{\nu_e}u_x^{-1},\qquad e:x\longrightarrow y.
 \label{eq:sm-filter-frames}
\end{equation}
For any global element $c\in G$, the simultaneous transformation $u_x\mapsto u_xc$ and $k\mapsto c^{-1}kc$ leaves $U_e(u,k)$ unchanged, while fixing the frame at one root vertex makes the map $(u,k)\mapsto U$ bijective onto $\mathscr F_x$. Each flat physical configuration thus has exactly $q$ coordinate representations, confirming $|\mathscr F_x|=q^V$. This $q$-fold multiplicity is purely a coordinate redundancy on the full edge Hilbert space rather than a gauge quotient.

In the replica sum $Z_R^{(2)}(\lambda)$, we introduce four sets of frame coordinates: $(u,k)$ and $(v,l)$ for the two base configurations, and $(u^{(3)},k^{(3)})$ and $(u^{(4)},k^{(4)})$ for the two swapped configurations. Summing over all four frame representations overcounts each physical configuration pair by $q^4$, which we compensate with a factor of $q^{-4}$. By definition of the swap, layers $3$ and $4$ match layers $1$ and $2$ on the exterior $\bar R$, and match layers $2$ and $1$ on the interior $R$. On any connected strip, equality of two flat connections forces their vertex frames to differ by a strip-wide right-multiplication constant, with the longitudinal twists related by the corresponding conjugation. We may therefore set $u^{(3)}=uc_i$ and $u^{(4)}=vd_i$ on each exterior strip, and $u^{(3)}=vc_i$ and $u^{(4)}=ud_i$ on each interior strip. Along the $j$th active boundary circle, consistency of the shared boundary vertices forces $v_x=u_xz_j$ for a seam constant $z_j\in G$, while matching the longitudinal winding edges requires the intertwining relation $kz_j=z_jl$. Passing from an exterior strip to an adjacent interior strip across interface $j$ updates the strip constants via $c_{j+1}=z_j^{-1}c_j$ and $d_{j+1}=z_jd_j$, whereas passing from an interior strip to an exterior strip multiplies by the inverse factors $z_j$ and $z_j^{-1}$. Periodic closure around the $n=2m_R$ alternating strips is therefore equivalent to restricting the tuple $\sigma=(k,l,z_1,\ldots,z_n)$ to the algebraic sector set $\Sigma_n$ defined by
\begin{equation}
\begin{gathered}
 kz_j=z_jl\quad(1\le j\le n),\\
 z_nz_{n-1}^{-1}\cdots z_2z_1^{-1}=1,\qquad
 z_n^{-1}z_{n-1}\cdots z_2^{-1}z_1=1.
\end{gathered}
\label{eq:sm-filter-constraints}
\end{equation}

An exterior--interior--exterior passage has the two-step update
\begin{equation}
 (c_j,d_j)\longmapsto(z_j^{-1}c_j,z_jd_j)
 \longmapsto(z_{j+1}z_j^{-1}c_j,\ z_{j+1}^{-1}z_jd_j).
 \label{eq:two-seam-updates}
\end{equation}

For a non-Abelian group $G$, neither alternating word condition in Eq.~\eqref{eq:sm-filter-constraints} may be dropped in general: if $a,c\in G$ do not commute, setting $k=l=1$ and $(z_1,z_2,z_3,z_4)=(1,a,c,a^{-1}c)$ satisfies $z_4z_3^{-1}z_2z_1^{-1}=1$ while yielding $z_4^{-1}z_3z_2^{-1}z_1=c^{-1}aca^{-1}\ne1$, and inverting each $z_j$ reverses which of the two products equals the identity. Conversely, given any tuple $\sigma\in\Sigma_n$ and base vertex frames obeying $v_x=u_xz_j$ on the active interfaces, choosing the initial strip constants $(c_1,d_1)\in G^2$ uniquely fixes all remaining $c_i,d_i$ and the swapped twists $k^{(3)},k^{(4)}$ via conjugation, reconstructing the same pair of swapped physical configurations in $q^2$ coordinate ways. Combining those $q^2$ choices with the $q^{-4}$ four-layer coordinate factor leaves a net prefactor of $q^{-2}$.

For each fixed sector $\sigma\in\Sigma_n$, the remaining frame degrees of freedom consist of one Haar-random variable $u_x$ at each of the $B_R$ active boundary vertices (where $v_x=u_xz_j$ is locked) and an independent Haar pair $(u_x,v_x)$ at each of the $V-B_R$ non-boundary vertices. Denoting the normalized product Haar measure over these $2V-B_R$ free group variables by $\mu_{R,\sigma}$, we arrive at the exact replica representation
\begin{equation}
\begin{gathered}
Z_R^{(2)}=q^{2V-B_R-2}\sum_{\sigma\in\Sigma_n}\Xi_{R,\sigma},\\
\Xi_{R,\sigma}=
\mathbb E_{\mu_{R,\sigma}}\prod_{e\ {\mathrm{physical}}}
 e^{-2\lambda[b(U_e(u,k))+b(U_e(v,l))]}.
\end{gathered}
\label{eq:sm-filter-reduced}
\end{equation}
Here the product extends over every physical edge of the refined multigraph, including both copies of all dormant and active seams. At the undeformed point $\lambda=0$, we have $\Xi_{R,\sigma}=1$ and $Z_0=q^V$, so Eq.~\eqref{eq:sm-filter-reduced} yields $P_R(0)=q^{-B_R-2}|\Sigma_{2m}|$. Comparing this expression with the exact fixed-point purity from Eq.~\eqref{sm:spectral-moment} gives
\begin{equation}
 P_R(0)=q^{-B_R-2}K_{2m}
       =q^{-B_R+2m-4}T_G(4-2m).
 \label{eq:replica-count-comparison}
\end{equation}
We thus obtain the exact cardinalities $K_n:=|\Sigma_n|$ for arbitrary finite groups without evaluating a filtered-spectrum ansatz:
\begin{equation}
\begin{gathered}
 K_{2m}:=|\Sigma_{2m}|=q^{2m-2}T_G(4-2m),\\
 K_2=q^2,\qquad K_4=q^2r_G,\qquad K_6=q^4T_G(-2).
\end{gathered}
\label{eq:sm-filter-counts}
\end{equation}
The integer $K_n$ counts algebraic replica tuples $(k,l,z_1,\ldots,z_n)\in\Sigma_n$ prior to any conjugation quotient, rather than the rank of a reduced density matrix.

Substituting Eq.~\eqref{eq:sm-filter-counts} into the four-region ratio at $\lambda=0$ immediately gives
\[
 -\log(K_2K_6/K_4^2)=\delta_G.
\]
By the Cauchy--Schwarz inequality applied to the sequences $d_a$ and $d_a^{-1}$ over $a\in\operatorname{Irr}D(G)$, we have $(\sum_a d_a^2)\allowbreak(\sum_a d_a^{-2})\allowbreak=q^2T_G(-2)\allowbreak\ge r_G^2$, with equality if and only if $d_a$ is constant across all sectors. Because any finite non-Abelian group possesses at least one sector with $d_a>1$ in addition to the trivial sector $d_0=1$, the inequality is strict and $\delta_G<0$.

\subsection{Uniform expansion and cancellation of individual interfaces}
To control the sector partition functions $\Xi_{R,\sigma}$ for $\lambda\ne0$, we decompose each physical-edge Boltzmann factor as $1+\zeta_e$, where $|\zeta_e|\le v_\lambda=e^{4|\lambda|}-1$. Because the reference measure $\mu_{R,\sigma}$ is a product over physical vertices, expectations of edge products factorize across vertex-disjoint edge sets. Expanding the product over all physical edges thus generates a hard-core polymer gas whose polymers $\gamma$ are nonempty connected subsets of physical edges, with two polymers deemed incompatible whenever they share at least one vertex. The activity of a polymer $\gamma$ satisfies
\begin{equation}
 a_\gamma=\mathbb E_{\mu_{R,\sigma}}\prod_{e\in\gamma}\zeta_e,
 \qquad |a_\gamma|\le v_\lambda^{|\gamma|}.
 \label{eq:sm-filter-activity}
\end{equation}
Any connected $k$-edge subgraph containing a given vertex $x$ can be traversed by a canonical Euler walk of length $2k$ on its doubled edge multigraph, so the number of polymers of size $|\gamma|=k$ pinned at $x$ is at most $\mathfrak d^{2k}$ (counting parallel bigon edges as distinct steps). Assigning each polymer the size weight $a(\gamma)=|V(\gamma)|\le2|\gamma|$, we verify the Koteck\'y--Preiss convergence criterion for the enhanced activities $|a_\gamma|e^{|\gamma|}$ across the entire disk $|\lambda|\le\lambda_{\rm CE}$:
\begin{equation}
 \sum_{\gamma\ni x}|a_\gamma|e^{|\gamma|+a(\gamma)}
 \le\sum_{k\ge1}\bigl(\mathfrak d^2 e^3 v_\lambda\bigr)^k\le1
 \quad\text{if }|\lambda|\le\lambda_{\rm CE}.
 \label{eq:sm-filter-kp}
\end{equation}
By Ueltschi's rooted cluster theorem~\cite[Theorem~1]{Ueltschi2004ClusterExpansion}, $\log\Xi_{R,\sigma}$ expands as an absolutely convergent series over connected polymer clusters $\mathcal K$:
\begin{equation}
 \log\Xi=\sum_{\mathcal K}c_{\mathcal K},\qquad
 \sum_{\mathcal K}|c_{\mathcal K}|e^{N(\mathcal K)}\le V.
 \label{eq:sm-filter-tail}
\end{equation}
Here $N(\mathcal K)=\sum_{\gamma\in\mathcal K}|\gamma|$ is the total edge count of the cluster including polymer multiplicities. To obtain the unrooted volume bound in Eq.~\eqref{eq:sm-filter-tail} explicitly, one multiplies Ueltschi's polymer-rooted estimate by the enhanced root activity and sums over all root polymers containing a fixed vertex $x$; each unrooted cluster $\mathcal K$ then appears with multiplicity $m_x(\mathcal K)$ equal to the number of its constituent polymers that touch $x$, and Eq.~\eqref{eq:sm-filter-kp} bounds that local sum by $1$. Summing over all $V$ vertices weights every nonempty cluster by $\sum_x m_x(\mathcal K)\ge1$, yielding Eq.~\eqref{eq:sm-filter-tail}. Restricting the sum to clusters with $N(\mathcal K)\ge s_*$ immediately bounds their combined absolute contribution by $Ve^{-s_*}$, with no dependence on the group order $q$ or the sector count $K_n$.

We now partition the connected clusters according to their geometric support on the torus. Call a cluster support \emph{bad} if it either intersects two or more distinct boundary circles from the common pool $\partial A_+\cup\partial B_+$ or contains a closed cycle of nonvanishing longitudinal winding around the torus; by definition of $W$ and $L$, every bad cluster has $N(\mathcal K)\ge\min(W,L)=s_*$. Every remaining \emph{good} support intersects at most one pool interface and carries an exact longitudinal cocycle $\nu_e=h_y-h_x$ for some integer potential $h_x$ on its vertices. On any good support, we perform the Haar-measure-preserving change of variables
\begin{equation}
 u'_x=u_xk^{h_x},\qquad v'_x=v_xl^{h_x}
 \label{eq:sm-filter-gauge}
\end{equation}
which removes $k$ and $l$ from the physical edge labels, leaving $U_e=u'_y(u'_x)^{-1}$ and $V_e=v'_y(v'_x)^{-1}$. If the good support meets a single active seam with intertwiner $z_j$, the relation $kz_j=z_jl$ implies $v'_x=u'_xz_j$ along that seam. Right-multiplying all second-layer frames $v'_x$ on the support by $z_j^{-1}$ then reduces the seam constraint to the universal identification $v''_x=u'_x$ while leaving every second-layer edge variable $V_e=v''_y(v''_x)^{-1}$ invariant. Because these frame shifts preserve the product Haar measure, the cluster coefficient $c_{\mathcal K}$ of any good support is strictly independent of the sector tuple $\sigma=(k,l,z_1,\ldots,z_n)$ and depends only on whether the single pool seam it touches is active in $R$. Crucially, this sector independence holds for arbitrary functions $b:G\to[0,1]$ without assuming $b(cg c^{-1})=b(g)$.

We apply this universal good/bad classification across all four regions $R\in\{A_+,\allowbreak B_+,\allowbreak U,\allowbreak I\}$ on the same refined multigraph (retaining all dormant and active duplicate edges in the polymer expansion). Let $\mathcal B$ be the sum of $c_{\mathcal K}$ over all good clusters evaluated in the reference measure with no active seams, and for each pool interface $j$ let $\mathcal H_j$ be the difference between the good-cluster sum with only interface $j$ active and $\mathcal B$. Both $\mathcal B$ and $\mathcal H_j$ are sector-independent analytical reference sums on the fixed graph. Grouping the cluster expansion of $\log\Xi_{R,\sigma}$ into good and bad terms gives
\begin{equation}
 \log\Xi_{R,\sigma}=\mathcal B+
 \sum_{j\in\partial R}\mathcal H_j+\epsilon_{R,\sigma},
 \qquad |\epsilon_{R,\sigma}|\le Ve^{-s_*}.
 \label{eq:sm-filter-local-log}
\end{equation}
Because $\lambda$ is real, each $\Xi_{R,\sigma}$ is strictly positive. Factoring out the sector-independent prefactor $\exp(\mathcal B+\sum_{j\in\partial R}\mathcal H_j)$ before summing over $\sigma\in\Sigma_n$ shows that the logarithmic sector average obeys
\begin{equation}
 \left|\log\left(K_n^{-1}\sum_{\sigma\in\Sigma_n}
 e^{\epsilon_{R,\sigma}}\right)\right|\le Ve^{-s_*}.
 \label{eq:sm-filter-sector-log}
\end{equation}
This inequality bounds the post-sum logarithm directly; it does not assume that individual replica sectors carry equal weights at finite volume.

In the high-order geometry, the active interface sets for $A_+,\allowbreak B_+,\allowbreak U,\allowbreak I$ are $\{1,\allowbreak 5\}$, $\{0,\allowbreak 2,\allowbreak 4,\allowbreak 6,\allowbreak 8,\allowbreak 9\}$, $\{0,\allowbreak 6,\allowbreak 8,\allowbreak 9\}$, and $\{1,\allowbreak 2,\allowbreak 4,\allowbreak 5\}$, so every pool interface $j$ enters the signed combination $A_++B_+-U-I$ with net multiplicity zero. When we assemble $\Delta_2^{\rm micro}\allowbreak=-\log(P_{A_+}P_{B_+}/P_UP_I)$ from Eq.~\eqref{eq:sm-filter-reduced}, the normalization factors $Z_\lambda$, the volume prefactors $q^{2V-2}$, the boundary-length powers $q^{-B_R}$, the bulk sum $\mathcal B$, and the single-seam terms $\mathcal H_j$ cancel out completely. The leading sector-count ratio $-\log(K_2K_6/K_4^2)$ reproduces $\delta_G$, while the four remainder terms from Eq.~\eqref{eq:sm-filter-sector-log} add up to at most $4Ve^{-s_*}$ in magnitude, completing the proof of Eq.~\eqref{eq:sm-filter-bound}.

\subsection{Changed reduced spectra and the \texorpdfstring{$S_3$}{S3} example}
To establish the finite-volume spectral variation in Eq.~\eqref{eq:sm-filter-spectrum-change}, assume the original straight-strip conditions $L,W\ge3$ with no two-edge cycles other than the installed seam bigons. Write $\mu_b=\mathbb E_{\rm Haar}[b]$ and $\sigma_b^2=\operatorname{Var}_{\rm Haar}(b)$. Within any fixed replica sector $\sigma$ at $\lambda=0$, define the total replica edge cost $C_R=\sum_e[b(U_e)+b(V_e)]$, and let $C_0$ be the corresponding sum in two independent normalization layers. Taking the first two $\lambda$-derivatives of $\log P_R(\lambda)$ at $\lambda=0$ yields
\begin{equation}
\begin{gathered}
 (\log P_R)'(0)=-2\bigl(\mathbb E C_R-\mathbb E C_0\bigr),\\
 (\log P_R)''(0)=4\bigl(\operatorname{Var}C_R-\operatorname{Var}C_0\bigr).
\end{gathered}
\label{eq:filter-variance-identity}
\end{equation}
By the single-edge good-support transformation, every physical edge in every sector has marginally Haar-distributed labels $U_e,V_e$, so each scalar cost has mean $\mu_b$ and $\mathbb E C_R=\mathbb E C_0$. Hence $(\log P_R)'(0)=0$. Under the no-extra-short-cycle hypothesis, any connected support consisting of at most two distinct geometric edges is good and forms a forest whose statistics are independent of $\sigma$. Vertex-disjoint supports factorize under the fixed-sector product measure, and their single-edge means are sector-independent. Integrating out the frame variable at a degree-one leaf vertex of a two-edge forest yields the conditional mean $2\mu_b$ for that leaf edge's cost $b(U_e)+b(V_e)$, regardless of the frame values at the other endpoint—both when the two replica frames at the leaf are independent and, by left/right Haar invariance, when they are locked by an active seam. Consequently, edge costs on distinct geometric edges have zero covariance in both the numerator and normalization measures. For any ordinary non-bigon edge, at least one endpoint lies off the active seam and carries independent layer frames, producing the exact same single-edge variance as in $C_0$. Likewise, a dormant bigon has both physical copies inside the same strip, so its within-layer duplicate covariance is identical in $C_R$ and $C_0$ and cancels in the difference.

A net variance difference arises exclusively at the $B_R$ active seam bigons. On an active bigon, the two duplicated physical edges share the same group label within each replica, and the active seam locks the two replicas together, forcing all four edge registers to take a single common Haar-random value $g$. The replica cost on that bigon is $4b(g)$ with variance $16\sigma_b^2$. In each uncoupled normalization layer, the two physical bigon copies still have equal labels. The two layer values $g_1,g_2$ are independent and Haar-distributed, giving cost $2b(g_1)+2b(g_2)$ with variance $8\sigma_b^2$. Each active bigon thus contributes
\begin{equation}
 4\left[16\sigma_b^2-8\sigma_b^2\right]=32\sigma_b^2.
 \label{eq:active-bigon-variance}
\end{equation}
Summing over the $B_R$ active bigons gives $32\sigma_b^2B_R$ in every sector $\sigma$; because neither the first nor the second moment depends on $\sigma$, averaging over $\Sigma_n$ introduces no inter-sector variance of the means, proving Eq.~\eqref{eq:sm-filter-spectrum-change}. Whenever $b$ is nonconstant ($\sigma_b^2>0$), $\log P_R(\lambda)$ has strictly positive second derivative at $\lambda=0$. Because $P_R(\lambda)$ is a ratio of finite exponential sums on a finite lattice, it is real-analytic in $\lambda$, ensuring that the reduced spectrum of $\rho_R(\lambda)$ genuinely varies with $\lambda$ in a neighborhood of $\lambda=0$ at every fixed finite volume (without implying monotonicity of $P_R(\lambda)$ across the full interval $|\lambda|\le\lambda_{\rm CE}$).

As a concrete illustration for $G=S_3$ ($q=6$), select a reflection $\tau\in S_3$ and set $b(g)=\mathbf1_{g\notin\{1,\tau\}}$. Up to overall normalization, this diagonal weight reproduces the positive $Q(k_1,0,0)$ deformation studied in Ref.~\cite{XuGarreRubioSchuch2022NonAbelianTransitions} under the identification $k_1=\lambda$. Because $\{1,\tau\}$ has cardinality $2$, the uniform Haar mean and variance are $\mu_b=2/3$ and $\sigma_b^2=2/9$, yielding the explicit finite-volume purity curvature $(\log P_R)''(0)=64B_R/9>0$. Meanwhile, the quantum dimensions $(1,1,2,3,3,2,2,2)$ of $D(S_3)$ give the exact replica sector counts $(K_2,K_4,K_6)=(36,288,4176)$ and the asymptotic deficit $\delta_{S_3}=\log(16/29)$, matching Eq.~\eqref{eq:microscopic-values}.

\subsection{A uniformly gapped local parent}
The replica estimate~\eqref{eq:sm-filter-bound} uses no Hamiltonian-gap assumption. We now construct a frustration-free local parent for $|\Psi_\lambda\rangle$ and establish its uniform spectral gap independently. We build upon Theorem~3.1 of Cui et al.~\cite{CuiEtAl2020QuantumDoubleCode}, which establishes for any finite-group quantum double on an ordinary lattice that all local zero-energy states on a contractible rectangular collar have identical reduced density matrices in the interior. Applying the polarization identity to this local indistinguishability property yields the first topological quantum order condition, TQO-1, while comparing the supports of reduced local and global ground-space projectors yields condition TQO-2, supplying the exact hypotheses of the Bravyi--Hastings--Michalakis (BHM) gap-stability theorem~\cite{BravyiHastingsMichalakis2010Stability}.

To lift the unrefined parent Hamiltonian $H_{\rm base}$ to our common bigon refinement, we initialize every duplicated edge register in the product ancilla state $|1\rangle$ penalized by $h_{\rm anc}=1-|1\rangle\langle1|$, and apply a depth-one circuit $C$ consisting of disjoint controlled-multiplication gates $|g,h\rangle\mapsto|g,hg\rangle$ on each bigon pair. The conjugated Hamiltonian $H_{\rm ref}=C(H_{\rm base}+H_{\rm anc})C^\dagger$ is a finite-range commuting-projector sum whose ground space coincides with the refined quantum-double ground space; in particular, $C$ maps the unrefined reference state tensored with ancillas directly onto $|\Omega_\perp\rangle$. Note that we only require this exact equality of ground subspaces, rather than a term-by-term identification between $H_{\rm ref}$ and the canonical refined star and plaquette projectors.

Let $P$ and $P_B$ denote the global and local ground-space projectors of $H_{\rm base}+H_{\rm anc}$, and let $P'=CPC^\dagger$ and $P'_B$ denote those of $H_{\rm ref}$. Because $C$ has depth one, conjugation by $C$ expands operator supports by a fixed graph radius $r$. If a local operator $O_A$ supported on $A$ satisfies $O_AP'=0$, then $Q=C^\dagger O_AC$ is supported on the $r$-collar $A^{+r}$ and satisfies $QP=0$. By condition TQO-2 for $H_{\rm base}+H_{\rm anc}$, $Q$ must also annihilate the local ground projector $P_B$ on an appropriate disk buffer $B\supset A^{+r}$. Because every interaction term of $H_{\rm base}+H_{\rm anc}$ supported inside $B$ maps under $C$ to a term of $H_{\rm ref}$ supported inside $B^{+r}$, the local ground projectors obey the operator inequality
\begin{equation}
 P'_{B^{+r}}\le CP_BC^\dagger.
 \label{eq:sm-filter-tqo-transfer}
\end{equation}
Multiplying on the left by $O_A$ gives $O_AP'_{B^{+r}}=0$; since global ground states are automatically local zero-energy states, the opposite kernel inclusion is immediate, establishing TQO-2 for $H_{\rm ref}$. The companion condition TQO-1 transfers identically under $A\mapsto A^{+r}$, and blocking the lattice by a fixed constant factor restores the standard unit-collar convention with the TQO conditions valid up to a fixed fraction of $\min(N_x,L)$ on bounded-aspect-ratio tori at fixed $G$.

Expanding $H_{\rm ref}=\sum_X h_X$ in terms of its local commuting projectors $h_X$ supported on edge sets $X$ of cardinality $|X|\le s_0$, we define the deformed frustration-free parent Hamiltonian $H(\lambda)=\sum_X h_X(\lambda)$ via an invertible congruence of each local term:
\begin{equation}
\begin{gathered}
 h_X(\lambda)=F_{\lambda,X}^{-1}h_XF_{\lambda,X}^{-1},\qquad
 \ker H(\lambda)=F_\lambda\ker H_{\rm ref},\\
 \|h_X(\lambda)-h_X\|\le e^{2|\lambda|s_0}-1,\qquad |X|\le s_0.
\end{gathered}
\label{eq:sm-filter-parent}
\end{equation}
Here $F_{\lambda,X}=\bigotimes_{e\in X}f_\lambda$ is the restriction of $F_\lambda$ to $X$. Because each $h_X(\lambda)$ is positive semidefinite, a vector $|\psi\rangle$ belongs to $\ker H(\lambda)$ if and only if $h_X(\lambda)|\psi\rangle=0$ for all $X$; since the invertible filter factors outside $X$ commute with $h_X$, this condition is equivalent to $h_XF_\lambda^{-1}|\psi\rangle=0$ for all $X$, proving the exact kernel identity $\ker H(\lambda)=F_\lambda\ker H_{\rm ref}$ (and hence $H(\lambda)|\Psi_\lambda\rangle=0$). Writing $B=F_{\lambda,X}^{-1}$ and using $0\le b(g)\le1$, the algebraic identity $Bh_XB-h_X=(B-I)h_XB+h_X(B-I)$ together with $\|B\|\le e^{|\lambda|s_0}$ and $\|B-I\|\le e^{|\lambda|s_0}-1$ yields the stated norm bound $\|h_X(\lambda)-h_X\|\le e^{2|\lambda|s_0}-1$, where the maximum support size $s_0$, interaction range, and vertex incidence are fixed by the local lattice geometry independently of $b$ and the torus dimensions.

To verify that the gap threshold can be chosen uniformly over all functions $b:G\to[0,1]$, fix the coarse-grained blocking, maximum interaction-box diameter $r_0$, exponential decay parameter $\mu>0$, and local term-incidence bound $M_0$, all of which depend only on the reference graph. The BHM local perturbation strength $J(\lambda)$ is then bounded above by
\begin{equation}
 J(\lambda)\le\kappa\bigl(e^{2s_0|\lambda|}-1\bigr),
 \qquad \kappa=M_0e^{\mu r_0},
 \label{eq:filter-local-strength}
\end{equation}
after absorbing any fixed local decomposition constants into $\kappa$. Let $J_*(G)>0$ be a strictly positive constant lying below both the gap-stability threshold and the lowest-band isolation threshold of the BHM theorem~\cite{BravyiHastingsMichalakis2010Stability} for the fixed group $G$. Defining the $b$-independent radius
\begin{equation}
 \lambda_{\rm gap}(G)=\frac{1}{2s_0}
 \log\left(1+\frac{J_*(G)}{\kappa}\right)>0
 \label{eq:filter-gap-radius}
\end{equation}
ensures $J(\lambda)\le J_*(G)$ for every $|\lambda|\le\lambda_{\rm gap}(G)$ and every $b:G\to[0,1]$. On all sufficiently large bounded-aspect-ratio tori, applying BHM stability along the linear interpolation $H_{\rm ref}+t[H(\lambda)-H_{\rm ref}]$ for $t\in[0,1]$ guarantees that the isolated lowest spectral band varies continuously without closing the gap above it, preserving the rank of its spectral projector equal to $\dim\ker H_{\rm ref}$. At $t=1$, the exact kernel relation $\ker H(\lambda)=F_\lambda\ker H_{\rm ref}$ provides $\dim\ker H_{\rm ref}$ linearly independent zero-energy states, forcing the entire isolated lowest band of $H(\lambda)$ to sit at exact zero energy with a uniform spectral gap above the ground space. Combining this gap stability on $|\lambda|\le\lambda_{\rm gap}(G)$ with the cluster bound on $|\lambda|\le\lambda_{\rm CE}$ completes the proof of Theorem~\ref{thm:filtered}.

%% file: sections/discussion.tex
Universal spatial strong subadditivity of the canonical topological entropy functional across pure torus ground states and regular subregions depends sharply on the R\'enyi order $\alpha$. At the von Neumann point $\alpha=1$, homological constraints combined with the dimension-weighted modular uncertainty relation enforce the inequality for every finite unitary modular tensor category. For finite orders $\alpha>0$ with $\alpha\ne1$, however, superpositions of sectors with distinct quantum dimensions induce strict curvature in the shared-flux functional as a function of the parallel-annulus count, generating explicit violations in every nonpointed theory. Within the low-order regime $0<\alpha<1$, all-state, all-region strong subadditivity holds if and only if the category is pointed. By contrast, every nontrivial category admits a fixed pure state and region pair that violate the inequality for all finite orders $\alpha\ge2$, with a crossing-annulus pair sufficing in the pointed case. Universal validity for $0<\alpha<1$ therefore distinguishes pointed from nonpointed anyon structures, whereas the mere presence of a high-order violation does not.

Because this classification concerns the constrained topological functional rather than ordinary reduced-state entropies, generic R\'enyi no-go theorems do not rule out its positive regimes~\cite{LindenMosonyiWinter2013RenyiInequalities}. On a disk, the canonical topological contribution remains insensitive to the R\'enyi index~\cite{FlammiaHammaHughesWen2009TopologicalRenyi}, whereas noncontractible cuts on the torus couple the functional to the global logical state. Within this torus setting, Lo and Chang announced the strong subadditivity property~\cite{LoChang2024TorusKnots} and reduced arbitrary region pairs to ribbon configurations~\cite{LoChang2025TorusSSA}. On the analytic side, the fusion-Fourier uncertainty relation of Liu, Palcoux, and Wu~\cite{LiuPalcouxWu2021Fusion} provides the required modular inequality. What establishes the positive regimes here is pairing that existing modular bound with an independent homological analysis of the region unions and intersections. The R\'enyi classification and its explicit spatial counterexamples reveal how shared flux and the entropy order together control when the inequality survives.

The microscopic constructions realize the shared-flux violations in ordinary physical-edge entropies. For each finite non-Abelian group, one specified pure quantum-double ground state on a common refined lattice gives a negative deficit on the low-order pair for $0<\alpha<1$ and on the high-order pair for every finite $\alpha>1$. All reductions use the same state and physical-edge factors. For each bipartition, the computational coordinate changes act separately on its two sides. Applying the same real diagonal filter generated by $b:G\to[0,1]$ to every physical edge extends the R\'enyi-two witness beyond the fixed point. At fixed group, the replica estimate controls the entropy deficit uniformly in $b$ on a common small parameter interval. A separate local-parent construction uses the reference model's topological quantum order to establish a uniform excitation gap on sufficiently large bounded-aspect-ratio tori in that interval. For nonconstant $b$, the regional spectra vary near zero at each fixed finite volume in the stated straight-strip geometry. The R\'enyi-two deficit tends to the strictly negative fixed-point value when $N_xL e^{-\min(W,L)}\to0$. The two proof chains therefore give an explicit gapped family with changing entanglement spectra and a controlled spatial violation.

Several natural questions arise when the logical state space or the ambient spatial topology is varied. Punctures and additional handles alter the flux data and admissible cut geometry, while other state preparations can change the sector probabilities. Within a fixed pointed category, sharpening the equal-order Fourier bound would refine the high-order part of the classification. A parallel microscopic question is how far the filtered-state analysis extends across broader families of wavefunctions away from the fixed point. For each of these directions, the present results establish the baseline state, geometric, and spectral structure.

%% file: sections/appendix-weighted.tex
\section{Weighted Fourier normalization}\label{app:uncertainty}
\[L=\log D.\]

The weighted uncertainty relation is an application of Theorem~4.9 of Liu, Palcoux and Wu~\cite{LiuPalcouxWu2021Fusion}. For a fusion matrix $N_a$, $N_a d=d_a d$ with $d>0$, so $\operatorname{diag}(d)^{-1}N_a\operatorname{diag}(d)$ has nonnegative entries and row sums $d_a$. Every eigenvalue thus has modulus at most $d_a$, without requiring irreducibility. Verlinde eigenvalues $S_{ab}/S_{0b}$ and $S_{0b}=d_b/D$ give
\begin{equation}
 |S_{ab}|\le\frac{d_ad_b}{D}.
 \label{eq:modular-entry-bound}
\end{equation}
For finite $r\ge1$, use $\|v\|_{r,\mu}=(\sum_a\mu_a|v_a|^r)^{1/r}$ with $\mu_a=d_a^2$, and the supremum norm at $r=\infty$. For normalized $\psi$, set
\begin{equation}
 f_a=\frac{\psi_a}{d_a},\qquad
 (\mathcal U f)_b=\frac1{d_b}\sum_a S_{ba}d_af_a
 =\sum_a\mu_a\frac{S_{ba}}{d_bd_a}f_a.
 \label{eq:weighted-transform}
\end{equation}
The kernel bound above and unitarity of $S$ give
\begin{equation}
 \|\mathcal U\|_{2\to2}=1,\qquad
 \|\mathcal U\|_{1\to\infty}\le D^{-1},\qquad
 \|f\|_{2,\mu}=\|\mathcal Uf\|_{2,\mu}=1.
 \label{eq:weighted-endpoints}
\end{equation}
Riesz--Thorin interpolation gives, for $0\le t<1$,
\begin{equation}
 \log\|\mathcal Uf\|_{s(t),\mu}-\log\|f\|_{r(t),\mu}\le-tL,
 \qquad r(t)=\frac2{1+t},\quad s(t)=\frac2{1-t}.
 \label{eq:interpolation-log}
\end{equation}
Both sides vanish at $t=0$. For $\|v\|_{2,\mu}=1$, $d\log\|v\|_{r,\mu}/dr|_{r=2}=\tfrac14\sum_a\mu_a|v_a|^2\log|v_a|^2$. With $r'(0)=-2$ and $s'(0)=2$, right differentiation gives
\begin{equation}
 H_\mu(f)+H_\mu(\mathcal Uf)\ge2L,
 \qquad H_\mu(v):=-\sum_a\mu_a|v_a|^2\log|v_a|^2.
 \label{eq:weighted-entropy-result}
\end{equation}
Because $\mu_a|f_a|^2=p_a$ and $\mu_b|(\mathcal Uf)_b|^2=q_b$, these entropies are $H(p)+2\ell_p$ and $H(q)+2\ell_q$. The vacuum saturates the result; zeros use $0\log0=0$. This interpolation method follows the classical entropic uncertainty approach~\cite{Hirschman1957,BialynickiBirulaMycielski1975}, its finite-dimensional and noncommutative formulations~\cite{MaassenUffink1988,JiangLiuWu2016Uncertainty}, and the review~\cite{ColesBertaTomamichelWehner2017Review}.

In the modular-category realization of the canonical fusion bialgebra, the two commutative algebras have minimal-projection traces $d_a^2$ and $d_b^2/D^2$. The normalized element and its Fourier transform have entries $\psi_a/d_a$ and $D(S\psi)_b/d_b$. Their algebraic entropies sum to
\begin{equation}
 H(p)+2\ell_p+H(q)+2\ell_q-2L.
 \label{eq:LPW-trace-dictionary}
\end{equation}
Its nonnegativity is exactly Eq.~\eqref{eq:weighted-uncertainty} in our normalization.

\paragraph{Normalization relative to the torus formula.}
Equation~(A.42) of Lo and Chang~\cite{LoChang2025TorusSSA} gives the integer-index factor $Z_{\rm top}(n)=\sum_a p_a^n(D/d_a)^{2m(n-1)}$. Its negative logarithmic derivative under the explicitly chosen finite-sum continuation at $n=1$ yields $-2mL+H(p)+2m\ell_p$: one Shannon term for the shared flux, not one per annulus. Their Eq.~(V.13) is half Eq.~\eqref{eq:weighted-uncertainty} in these conventions. The Liu--Palcoux--Wu trace dictionary supplies that input; the arbitrary-subsurface theorem additionally needs the geometric exhaustion above. No standalone density operator for $Z_{\rm top}$ is assumed.

%% file: torus-topological-entropy.bbl
\providecommand{\href}[2]{#2}\begingroup\raggedright\begin{thebibliography}{10}

\bibitem{KitaevPreskill2006}
A.~Kitaev and J.~Preskill, \emph{Topological entanglement entropy},
  \href{https://doi.org/10.1103/PhysRevLett.96.110404}{\emph{Phys. Rev. Lett.}
  {\bfseries 96} (2006) 110404}
  [\href{https://arxiv.org/abs/hep-th/0510092}{{\ttfamily hep-th/0510092}}].

\bibitem{LevinWen2006}
M.~Levin and X.-G.~Wen, \emph{Detecting topological order in a ground state
  wave function},
  \href{https://doi.org/10.1103/PhysRevLett.96.110405}{\emph{Phys. Rev. Lett.}
  {\bfseries 96} (2006) 110405}
  [\href{https://arxiv.org/abs/cond-mat/0510613}{{\ttfamily
  cond-mat/0510613}}].

\bibitem{LiebRuskai1973}
E.H.~Lieb and M.B.~Ruskai, \emph{Proof of the strong subadditivity of
  quantum-mechanical entropy},
  \href{https://doi.org/10.1063/1.1666274}{\emph{J. Math. Phys.} {\bfseries 14}
  (1973) 1938}.

\bibitem{LiHaldane2008}
H.~Li and F.D.M.~Haldane, \emph{Entanglement spectrum as a generalization of
  entanglement entropy: Identification of topological order in non-{Abelian}
  fractional quantum {Hall} effect states},
  \href{https://doi.org/10.1103/PhysRevLett.101.010504}{\emph{Phys. Rev. Lett.}
  {\bfseries 101} (2008) 010504}.

\bibitem{DongEtAl2008}
S.~Dong, E.~Fradkin, R.G.~Leigh and S.~Nowling, \emph{Topological entanglement
  entropy in {Chern--Simons} theories and quantum {Hall} fluids},
  \href{https://doi.org/10.1088/1126-6708/2008/05/016}{\emph{J. High Energy
  Phys.} {\bfseries 05} (2008) 016}
  [\href{https://arxiv.org/abs/0802.3231}{{\ttfamily 0802.3231}}].

\bibitem{WenMatsuuraRyu2016Edge}
X.~Wen, S.~Matsuura and S.~Ryu, \emph{Edge theory approach to topological
  entanglement entropy, mutual information, and entanglement negativity in
  {Chern-Simons} theories},
  \href{https://doi.org/10.1103/physrevb.93.245140}{\emph{Phys. Rev. B}
  {\bfseries 93} (2016) 245140}
  [\href{https://arxiv.org/abs/1603.08534}{{\ttfamily 1603.08534}}].

\bibitem{BondersonKnappPatel2017}
P.~Bonderson, C.~Knapp and K.~Patel, \emph{Anyonic entanglement and topological
  entanglement entropy},
  \href{https://doi.org/10.1016/j.aop.2017.07.018}{\emph{Ann. Phys.} {\bfseries
  385} (2017) 399} [\href{https://arxiv.org/abs/1706.09420}{{\ttfamily
  1706.09420}}].

\bibitem{LoChang2024TorusKnots}
C.-Y.~Lo and P.-Y.~Chang, \emph{Topological entanglement entropy for torus-knot
  bipartitions and the {Verlinde}-like formulas},
  \href{https://doi.org/10.1007/jhep02(2024)117}{\emph{J. High Energy Phys.}
  {\bfseries 02} (2024) 117}
  [\href{https://arxiv.org/abs/2312.08348}{{\ttfamily 2312.08348}}].

\bibitem{LindenMosonyiWinter2013RenyiInequalities}
N.~Linden, M.~Mosonyi and A.~Winter, \emph{The structure of {R{\'e}nyi}
  entropic inequalities},
  \href{https://doi.org/10.1098/rspa.2012.0737}{\emph{Proc. R. Soc. A}
  {\bfseries 469} (2013) 20120737}
  [\href{https://arxiv.org/abs/1212.0248}{{\ttfamily 1212.0248}}].

\bibitem{LoChang2025TorusSSA}
C.-Y.~Lo and P.-Y.~Chang, \emph{Towards a topological proof of the strong
  subadditivity},  \href{https://arxiv.org/abs/2411.05077}{{\ttfamily
  2411.05077}}.

\bibitem{LiuPalcouxWu2021Fusion}
Z.~Liu, S.~Palcoux and J.~Wu, \emph{Fusion bialgebras and {Fourier} analysis},
  \href{https://doi.org/10.1016/j.aim.2021.107905}{\emph{Adv. Math.} {\bfseries
  390} (2021) 107905} [\href{https://arxiv.org/abs/1910.12059}{{\ttfamily
  1910.12059}}].

\bibitem{JianKimQi2015LRMI}
C.-M.~Jian, I.H.~Kim and X.-L.~Qi, \emph{Long-range mutual information and
  topological uncertainty principle},
  \href{https://arxiv.org/abs/1508.07006}{{\ttfamily 1508.07006}}.

\bibitem{ShiKatoKim2020}
B.~Shi, K.~Kato and I.H.~Kim, \emph{Fusion rules from entanglement},
  \href{https://doi.org/10.1016/j.aop.2020.168164}{\emph{Ann. Phys.} {\bfseries
  418} (2020) 168164} [\href{https://arxiv.org/abs/1906.09376}{{\ttfamily
  1906.09376}}].

\bibitem{ShiHuangMcGreevy2025TorusTrick}
B.~Shi, J.-L.~Huang and J.~McGreevy, \emph{Remote detectability from
  entanglement bootstrap {I}: {Kirby}’s torus trick},
  \href{https://doi.org/10.21468/scipostphys.18.4.126}{\emph{SciPost Phys.}
  {\bfseries 18} (2025) 126}
  [\href{https://arxiv.org/abs/2301.07119}{{\ttfamily 2301.07119}}].

\bibitem{Witten1989Jones}
E.~Witten, \emph{Quantum field theory and the {Jones} polynomial},
  \href{https://doi.org/10.1007/bf01217730}{\emph{Commun. Math. Phys.}
  {\bfseries 121} (1989) 351}.

\bibitem{ReshetikhinTuraev1991Invariants}
N.~Reshetikhin and V.G.~Turaev, \emph{Invariants of 3-manifolds via link
  polynomials and quantum groups},
  \href{https://doi.org/10.1007/bf01239527}{\emph{Invent. Math.} {\bfseries
  103} (1991) 547}.

\bibitem{FlammiaHammaHughesWen2009TopologicalRenyi}
S.T.~Flammia, A.~Hamma, T.L.~Hughes and X.-G.~Wen, \emph{Topological
  entanglement {R{\'e}nyi} entropy and reduced density matrix structure},
  \href{https://doi.org/10.1103/physrevlett.103.261601}{\emph{Phys. Rev. Lett.}
  {\bfseries 103} (2009) 261601}
  [\href{https://arxiv.org/abs/0909.3305}{{\ttfamily 0909.3305}}].

\bibitem{ZhangEtAl2012MES}
Y.~Zhang, T.~Grover, A.~Turner, M.~Oshikawa and A.~Vishwanath,
  \emph{Quasiparticle statistics and braiding from ground-state entanglement},
  \href{https://doi.org/10.1103/physrevb.85.235151}{\emph{Phys. Rev. B}
  {\bfseries 85} (2012) 235151}
  [\href{https://arxiv.org/abs/1111.2342}{{\ttfamily 1111.2342}}].

\bibitem{BertaSeshadreesanWilde2015RenyiCMI}
M.~Berta, K.P.~Seshadreesan and M.M.~Wilde, \emph{R{\'e}nyi generalizations of
  the conditional quantum mutual information},
  \href{https://doi.org/10.1063/1.4908102}{\emph{J. Math. Phys.} {\bfseries 56}
  (2015) 022205} [\href{https://arxiv.org/abs/1403.6102}{{\ttfamily
  1403.6102}}].

\bibitem{Kitaev2006}
A.~Kitaev, \emph{Anyons in an exactly solved model and beyond},
  \href{https://doi.org/10.1016/j.aop.2005.10.005}{\emph{Ann. Phys.} {\bfseries
  321} (2006) 2} [\href{https://arxiv.org/abs/cond-mat/0506438}{{\ttfamily
  cond-mat/0506438}}].

\bibitem{BialynickiBirula2006RenyiUncertainty}
I.~Bialynicki-Birula, \emph{Formulation of the uncertainty relations in terms
  of the {R{\'e}nyi} entropies},
  \href{https://doi.org/10.1103/physreva.74.052101}{\emph{Phys. Rev. A}
  {\bfseries 74} (2006) 052101}
  [\href{https://arxiv.org/abs/quant-ph/0608116}{{\ttfamily
  quant-ph/0608116}}].

\bibitem{PuchalaRudnickiZyczkowski2013Majorization}
Z.~Pucha{\l}a, {\L}.~Rudnicki and K.~{\.Z}yczkowski, \emph{Majorization
  entropic uncertainty relations},
  \href{https://doi.org/10.1088/1751-8113/46/27/272002}{\emph{J. Phys. A: Math.
  Theor.} {\bfseries 46} (2013) 272002}
  [\href{https://arxiv.org/abs/1304.7755}{{\ttfamily 1304.7755}}].

\bibitem{AbdelkhalekEtAl2015Optimality}
K.~Abdelkhalek, R.~Schwonnek, H.~Maassen, F.~Furrer, J.~Duhme, P.~Raynal
  et~al., \emph{Optimality of entropic uncertainty relations},
  \href{https://doi.org/10.1142/s0219749915500458}{\emph{Int. J. Quantum Inf.}
  {\bfseries 13} (2015) 1550045}
  [\href{https://arxiv.org/abs/1509.00398}{{\ttfamily 1509.00398}}].

\bibitem{ZozorBosykPortesi2013QubitRenyi}
S.~Zozor, G.M.~Bosyk and M.~Portesi, \emph{On a generalized entropic
  uncertainty relation in the case of the qubit},
  \href{https://doi.org/10.1088/1751-8113/46/46/465301}{\emph{J. Phys. A: Math.
  Theor.} {\bfseries 46} (2013) 465301}
  [\href{https://arxiv.org/abs/1306.0409}{{\ttfamily 1306.0409}}].

\bibitem{Kim2013}
I.H.~Kim, \emph{Long-range entanglement is necessary for a topological storage
  of quantum information},
  \href{https://doi.org/10.1103/physrevlett.111.080503}{\emph{Phys. Rev. Lett.}
  {\bfseries 111} (2013) 080503}
  [\href{https://arxiv.org/abs/1304.3925}{{\ttfamily 1304.3925}}].

\bibitem{BaoCaoWalter2015EtAl}
N.~Bao, C.~Cao, M.~Walter and Z.~Wang, \emph{Holographic entropy inequalities
  and gapped phases of matter},
  \href{https://doi.org/10.1007/jhep09(2015)203}{\emph{J. High Energy Phys.}
  {\bfseries 09} (2015) 203}
  [\href{https://arxiv.org/abs/1507.05650}{{\ttfamily 1507.05650}}].

\bibitem{NaskarSamal2026}
J.~Naskar and S.S.~Samal, \emph{Topological entanglement entropy meets
  holographic entropy inequalities},
  \href{https://doi.org/10.1103/hvf8-jcff}{\emph{Phys. Rev. D} {\bfseries 114}
  (2026) 026025} [\href{https://arxiv.org/abs/2412.05484}{{\ttfamily
  2412.05484}}].

\bibitem{KatoFurrerMurao2016Information}
K.~Kato, F.~Furrer and M.~Murao, \emph{Information-theoretical analysis of
  topological entanglement entropy and multipartite correlations},
  \href{https://doi.org/10.1103/physreva.93.022317}{\emph{Phys. Rev. A}
  {\bfseries 93} (2016) 022317}
  [\href{https://arxiv.org/abs/1505.01917}{{\ttfamily 1505.01917}}].

\bibitem{KatoBrandao2019EdgeLocality}
K.~Kato and F.G.S.L.~Brand{\~a}o, \emph{Locality of edge states and
  entanglement spectrum from strong subadditivity},
  \href{https://doi.org/10.1103/PhysRevB.99.195124}{\emph{Phys. Rev. B}
  {\bfseries 99} (2019) 195124}.

\bibitem{Shi2019InformationConvex}
B.~Shi, \emph{Seeing topological entanglement through the information convex},
  \href{https://doi.org/10.1103/physrevresearch.1.033048}{\emph{Phys. Rev.
  Research} {\bfseries 1} (2019) 033048}
  [\href{https://arxiv.org/abs/1810.01986}{{\ttfamily 1810.01986}}].

\bibitem{JiangLiuWu2016Uncertainty}
C.~Jiang, Z.~Liu and J.~Wu, \emph{Noncommutative uncertainty principles},
  \href{https://doi.org/10.1016/j.jfa.2015.08.007}{\emph{J. Funct. Anal.}
  {\bfseries 270} (2016) 264}
  [\href{https://arxiv.org/abs/1408.1165}{{\ttfamily 1408.1165}}].

\bibitem{MaassenUffink1988}
H.~Maassen and J.B.M.~Uffink, \emph{Generalized entropic uncertainty
  relations}, \href{https://doi.org/10.1103/physrevlett.60.1103}{\emph{Phys.
  Rev. Lett.} {\bfseries 60} (1988) 1103}.

\bibitem{Kitaev2003FaultTolerantAnyons}
{A. Yu. Kitaev}, \emph{Fault-tolerant quantum computation by anyons},
  \href{https://doi.org/10.1016/s0003-4916(02)00018-0}{\emph{Ann. Phys.}
  {\bfseries 303} (2003) 2}.

\bibitem{LevinWen2005StringNet}
M.A.~Levin and X.-G.~Wen, \emph{String-net condensation: A physical mechanism
  for topological phases},
  \href{https://doi.org/10.1103/physrevb.71.045110}{\emph{Phys. Rev. B}
  {\bfseries 71} (2005) 045110}.

\bibitem{BuerschaperAguado2009Mapping}
O.~Buerschaper and M.~Aguado, \emph{Mapping {Kitaev}'s quantum double lattice
  models to {Levin} and {Wen}'s string-net models},
  \href{https://doi.org/10.1103/physrevb.80.155136}{\emph{Phys. Rev. B}
  {\bfseries 80} (2009) 155136}.

\bibitem{SchuchCiracPerezGarcia2010PEPS}
N.~Schuch, I.~Cirac and D.~P{\'e}rez-Garc{\'i}a, \emph{{PEPS} as ground states:
  Degeneracy and topology},
  \href{https://doi.org/10.1016/j.aop.2010.05.008}{\emph{Ann. Phys.} {\bfseries
  325} (2010) 2153}.

\bibitem{HammaIonicioiuZanardi2005Kitaev}
A.~Hamma, R.~Ionicioiu and P.~Zanardi, \emph{Ground state entanglement and
  geometric entropy in the {Kitaev} model},
  \href{https://doi.org/10.1016/j.physleta.2005.01.060}{\emph{Phys. Lett. A}
  {\bfseries 337} (2005) 22}
  [\href{https://arxiv.org/abs/quant-ph/0406202}{{\ttfamily
  quant-ph/0406202}}].

\bibitem{LuoHuWu2016TorusEntanglement}
Z.-X.~Luo, Y.-T.~Hu and Y.-S.~Wu, \emph{Quantum entanglement in topological
  phases on a torus},
  \href{https://doi.org/10.1103/physrevb.94.075126}{\emph{Phys. Rev. B}
  {\bfseries 94} (2016) 075126}.

\bibitem{Liu2024QuantumDimensions}
S.~Liu, \emph{Anyon quantum dimensions from an arbitrary ground state wave
  function}, \href{https://doi.org/10.1038/s41467-024-47856-7}{\emph{Nat.
  Commun.} {\bfseries 15} (2024) 5134}
  [\href{https://arxiv.org/abs/2304.13235}{{\ttfamily 2304.13235}}].

\bibitem{CasiniHuertaRosabal2014GaugeEntropy}
H.~Casini, M.~Huerta and J.A.~Rosabal, \emph{{Remarks on entanglement entropy
  for gauge fields}},
  \href{https://doi.org/10.1103/physrevd.89.085012}{\emph{Phys. Rev. D}
  {\bfseries 89} (2014) 085012}.

\bibitem{Donnelly2012GaugeEntropy}
W.~Donnelly, \emph{{Decomposition of entanglement entropy in lattice gauge
  theory}}, \href{https://doi.org/10.1103/physrevd.85.085004}{\emph{Phys. Rev.
  D} {\bfseries 85} (2012) 085004}.

\bibitem{VanAcoleyenEtAl2016Distillation}
K.~Van~Acoleyen, N.~Bultinck, J.~Haegeman, M.~Marien, V.B.~Scholz and
  F.~Verstraete, \emph{Entanglement of distillation for lattice gauge
  theories}, \href{https://doi.org/10.1103/physrevlett.117.131602}{\emph{Phys.
  Rev. Lett.} {\bfseries 117} (2016) 131602}.

\bibitem{KawagoeEtAl2024LevinWenGauge}
K.~Kawagoe, C.~Jones, S.~Sanford, D.~Green and D.~Penneys, \emph{{Levin-Wen} is
  a gauge theory: Entanglement from topology},
  \href{https://doi.org/10.1007/s00220-024-05144-x}{\emph{Commun. Math. Phys.}
  {\bfseries 405} (2024) 266}.

\bibitem{CastelnovoChamon2008DeformedToric}
C.~Castelnovo and C.~Chamon, \emph{Quantum topological phase transition at the
  microscopic level},
  \href{https://doi.org/10.1103/physrevb.77.054433}{\emph{Phys. Rev. B}
  {\bfseries 77} (2008) 054433}.

\bibitem{PapanikolaouRamanFradkin2007FiniteCorrelation}
S.~Papanikolaou, K.S.~Raman and E.~Fradkin, \emph{Topological phases and
  topological entropy of two-dimensional systems with finite correlation
  length}, \href{https://doi.org/10.1103/physrevb.76.224421}{\emph{Phys. Rev.
  B} {\bfseries 76} (2007) 224421}.

\bibitem{HaegemanEtAl2015Shadows}
J.~Haegeman, V.~Zauner, N.~Schuch and F.~Verstraete, \emph{Shadows of anyons
  and the entanglement structure of topological phases},
  \href{https://doi.org/10.1038/ncomms9284}{\emph{Nat. Commun.} {\bfseries 6}
  (2015) 8284} [\href{https://arxiv.org/abs/1410.5443}{{\ttfamily 1410.5443}}].

\bibitem{Kim2012PerturbativeTEE}
I.H.~Kim, \emph{Perturbative analysis of topological entanglement entropy from
  conditional independence},
  \href{https://doi.org/10.1103/PhysRevB.86.245116}{\emph{Phys. Rev. B}
  {\bfseries 86} (2012) 245116}.

\bibitem{KimLevinLin2023EtAl}
I.H.~Kim, M.~Levin, T.-C.~Lin, D.~Ranard and B.~Shi, \emph{Universal lower
  bound on topological entanglement entropy},
  \href{https://doi.org/10.1103/physrevlett.131.166601}{\emph{Phys. Rev. Lett.}
  {\bfseries 131} (2023) 166601}
  [\href{https://arxiv.org/abs/2302.00689}{{\ttfamily 2302.00689}}].

\bibitem{Levin2024TEEBound}
M.~Levin, \emph{Physical proof of the topological entanglement entropy
  inequality}, \href{https://doi.org/10.1103/physrevb.110.165154}{\emph{Phys.
  Rev. B} {\bfseries 110} (2024) 165154}
  [\href{https://arxiv.org/abs/2408.04592}{{\ttfamily 2408.04592}}].

\bibitem{WilliamsonDuaCheng2019}
D.J.~Williamson, A.~Dua and M.~Cheng, \emph{Spurious topological entanglement
  entropy from subsystem symmetries},
  \href{https://doi.org/10.1103/physrevlett.122.140506}{\emph{Phys. Rev. Lett.}
  {\bfseries 122} (2019) 140506}
  [\href{https://arxiv.org/abs/1808.05221}{{\ttfamily 1808.05221}}].

\bibitem{KatoBrandao2020SpuriousBoundary}
K.~Kato and F.G.S.L.~Brand{\~a}o, \emph{Toy model of boundary states with
  spurious topological entanglement entropy},
  \href{https://doi.org/10.1103/PhysRevResearch.2.032005}{\emph{Phys. Rev.
  Research} {\bfseries 2} (2020) 032005(R)}.

\bibitem{ZouHaah2016}
L.~Zou and J.~Haah, \emph{Spurious long-range entanglement and replica
  correlation length},
  \href{https://doi.org/10.1103/physrevb.94.075151}{\emph{Phys. Rev. B}
  {\bfseries 94} (2016) 075151}
  [\href{https://arxiv.org/abs/1604.06101}{{\ttfamily 1604.06101}}].

\bibitem{OlivieroLeoneZhouHamma2022TopologicalPurity}
S.F.E.~Oliviero, L.~Leone, Y.~Zhou and A.~Hamma, \emph{Stability of topological
  purity under random local unitaries},
  \href{https://doi.org/10.21468/scipostphys.12.3.096}{\emph{SciPost Phys.}
  {\bfseries 12} (2022) 096}.

\bibitem{Ueltschi2004ClusterExpansion}
D.~Ueltschi, \emph{Cluster expansions and correlation functions},
  \href{https://doi.org/10.17323/1609-4514-2004-4-2-511-522}{\emph{Mosc. Math.
  J.} {\bfseries 4} (2004) 511}.

\bibitem{XuGarreRubioSchuch2022NonAbelianTransitions}
W.-T.~Xu, J.~Garre-Rubio and N.~Schuch, \emph{Complete characterization of
  non-{Abelian} topological phase transitions and detection of anyon splitting
  with projected entangled pair states},
  \href{https://doi.org/10.1103/physrevb.106.205139}{\emph{Phys. Rev. B}
  {\bfseries 106} (2022) 205139}.

\bibitem{CuiEtAl2020QuantumDoubleCode}
S.X.~Cui, D.~Ding, X.~Han, G.~Penington, D.~Ranard, B.C.~Rayhaun et~al.,
  \emph{Kitaev's quantum double model as an error correcting code},
  \href{https://doi.org/10.22331/q-2020-09-24-331}{\emph{Quantum} {\bfseries 4}
  (2020) 331}.

\bibitem{BravyiHastingsMichalakis2010Stability}
S.~Bravyi, M.B.~Hastings and S.~Michalakis, \emph{Topological quantum order:
  Stability under local perturbations},
  \href{https://doi.org/10.1063/1.3490195}{\emph{J. Math. Phys.} {\bfseries 51}
  (2010) 093512}.

\bibitem{Hirschman1957}
I.I.~Hirschman, \emph{A note on entropy},
  \href{https://doi.org/10.2307/2372390}{\emph{Am. J. Math.} {\bfseries 79}
  (1957) 152}.

\bibitem{BialynickiBirulaMycielski1975}
I.~Bia{\l}ynicki-Birula and J.~Mycielski, \emph{Uncertainty relations for
  information entropy in wave mechanics},
  \href{https://doi.org/10.1007/bf01608825}{\emph{Commun. Math. Phys.}
  {\bfseries 44} (1975) 129}.

\bibitem{ColesBertaTomamichelWehner2017Review}
P.J.~Coles, M.~Berta, M.~Tomamichel and S.~Wehner, \emph{Entropic uncertainty
  relations and their applications},
  \href{https://doi.org/10.1103/revmodphys.89.015002}{\emph{Rev. Mod. Phys.}
  {\bfseries 89} (2017) 015002}
  [\href{https://arxiv.org/abs/1511.04857}{{\ttfamily 1511.04857}}].

\end{thebibliography}\endgroup
